\documentclass[%
 aip,
 amsmath,amssymb,
 reprint,%
]{revtex4-1}

\usepackage{graphicx}
\usepackage{dcolumn}
\usepackage{bm}

\usepackage[utf8]{inputenc}
\usepackage[T1]{fontenc}
\usepackage{mathptmx}
\usepackage{etoolbox}
\usepackage{subcaption}
\usepackage{xcolor}
\usepackage{ulem}

\makeatletter
\def\@email#1#2{%
 \endgroup
 \patchcmd{\titleblock@produce}
  {\frontmatter@RRAPformat}
  {\frontmatter@RRAPformat{\produce@RRAP{*#1\href{mailto:#2}{#2}}}\frontmatter@RRAPformat}
  {}{}
}%
\makeatother
\begin{document}

\preprint{AIP/123-QED}

\title[]{Excitation of toroidal Alfvén eigenmode by energetic particles in DTT and effect of negative triangularity}
\author{Guangyu Wei}
\affiliation{Institute for Fusion Theory and Simulation and Department of Physics, Zhejiang University, Hangzhou 310027, China}
\affiliation{Center for Nonlinear Plasma Science and C.R. ENEA Frascati, Via E. Fermi 45, 00044 Frascati, Italy}
\author{Fulvio Zonca}
\affiliation{Center for Nonlinear Plasma Science and C.R. ENEA Frascati, Via E. Fermi 45, 00044 Frascati, Italy}
\affiliation{Institute for Fusion Theory and Simulation and Department of Physics, Zhejiang University, Hangzhou 310027, China}
\author{Matteo Valerio Falessi}
\affiliation{Center for Nonlinear Plasma Science and C.R. ENEA Frascati, Via E. Fermi 45, 00044 Frascati, Italy}
\affiliation{Istituto Nazionale di Fisica Nucleare (INFN), Sezione di Roma, Piazzale Aldo Moro 2, 00185 Roma, Italy}
\author{Zhiyong Qiu}\email{zqiu@ipp.ac.cn}
\affiliation{Key Laboratory of Frontier Physics in Controlled Nuclear Fusion and Institute of Plasma Physics, Chinese Academy of Sciences, Hefei 230031, China}
\affiliation{Center for Nonlinear Plasma Science and C.R. ENEA Frascati, Via E. Fermi 45, 00044 Frascati, Italy}

\date{\today}

\begin{abstract}
A linear gyrokinetic eigenvalue code is developed to study the stability of toroidal Alfvén eigenmode (TAE) in general axisymmetric toroidal geometry, with the self-consistent treatment of energetic particle drive and core plasma Landau damping in a non-perturbative way. The general particle responses of both circulating and trapped particles are  incorporated in the calculation by means of the action-angle approach, and, particularly, the finite Larmor radius and orbit width effects of energetic particles are fully taken into account. The ballooning-mode representation is adopted to solve the eigenmode equations in order to reduce the computational resource while obtaining a high resolution of the fine radial structure. Furthermore, the code is able to study the physics of wave-particle interaction in great detail, thanks to the development of systematic theory-based numerical diagnostics, including effective mode structure and phase space resonance structure. As an application of the code, we perform an in-depth study of the triangularity effect on TAE stability based on the reference equilibrium of the Divertor Tokamak Test facility. It is demonstrated that TAE growth rate can be affected by the triangularity through the modifications of geometric couplings, resonance condition, as well as mode frequency and mode structure. As a result, negative triangularity can either stabilize or destabilize the energetic particle driven TAE depending on the dominant mechanism. The relative importance of these mechanisms under different circumstances is systematically analyzed, providing clear physical insights. The overall effect of negative triangularity for a specific tokamak scenario can be assessed based on these studies. 

\end{abstract}

\maketitle

\section{Introduction}\label{Introduction}
In tokamak plasma, the variation of local shear Alfvén wave (SAW) frequency due to the plasma nonuniformities associated with equilibrium magnetic field and plasma profiles constitutes the SAW continuum, and SAW fluctuations with frequencies in the continuum will experience continuum damping via mode conversion to small-scale structures Laudau damped by, predominantly, electrons\cite{AHasegawaPRL1974,LChenPoF1974}. However, when symmetric breaking effects such as toroidal geometry and plasma compressibility are considered, frequency gaps form inside the continuum\cite{CZChengAP1985,MSChuPoF1992}, where discrete Alfvén eigenmodes (AEs), such as toroidal Alfvén eigenmode (TAE)\cite{CZChengAP1985,CZChengPoF1986} and beta-induced Alfvén eigenmode (BAE)\cite{WHeidbrinkPRL1993}, can exist and are free of significant continuum damping. In fusion plasma, energetic particles (EPs) generated from auxiliary heating and/or fusion reaction typically have characteristic velocities comparable with Alfvén speed, and thus may effectively excite AE instabilities through resonant wave-particle interactions\cite{WHeidbrinkPoP2008}. Among the various AEs, TAE is considered to be one of the most dangerous candidates for inducing considerable EP losses in future reactors\cite{WHeidbrinkPRL1991,KWongPRL1991,MGarciaMunozPRL2010}, which may lead to the degradation of plasma confinement and, potentially damage to the first wall components. Therefore, a comprehensive understanding of linear and nonlinear TAE physics is crucial for fusion plasma research\cite{LChenRMP2016,YTodoRMPP2019}. Up to now, the linear theory of TAE, including resonant excitation by EPs\cite{GFuPoFB1989,GFuPoFB1992,GFuPoFB1993,LChenPoP1994,FZoncaPoP1996}, continuum damping\cite{FZoncaPRL1992,FZoncaPoFB1993,HBerkPoFB1992,MRosenbluthPRL1992} and kinetic damping as realistic geometry and kinetic effects are accounted for\cite{RMettPoFB1992,GFuPoP1996,GFuPoP2006}, has been continuously developed, and qualitative understanding of TAE linear physics are well established. In particular, the general fishbone-like dispersion relation (GFLDR)\cite{LChenRMP2016,FZoncaPoP2014a,FZoncaPoP2014b} provides a unified theoretical framework for the description of SAW fluctuations, including TAE, in a wide range of spatial and temporal scales. Interested readers may refer to reference \citenum{LChenRMP2016} for a comprehensive review.

Due to the complexity of tokamak geometry and equilibrium profiles, numerical simulations are generally required to provide quantitative evaluation of AE instabilities in realistic experimental configuration. Particularly, the eigenvalue approach based on time-domain Fourier analysis is regarded as the most convenient way to investigate linear AE physics, for its fast computational speed and ability to capture all possible roots, including the stable modes. Over the last few decades, many eigenvalue codes\cite{CZChengPR1992,DBorbaJCP1999,PLauberJCP2007,JBaoNF2023} have been developed to calculate AE instabilities in the presence of EPs based on different models. In an axisymmetric toroidal geometry, the eigenmode problem in the configuration space is intrinsically a 2D problem for a given toroidal mode number $n$. In most existing codes, the governing equations are solved using finite difference or finite element discretization in the radial direction combined with Fourier decomposition in the poloidal direction, which transforms the problem into a matrix eigenvalue formulation. This approach provides straightforward implementation and direct access to the global mode structures. However, the computational cost may increase substantially for high-$n$ modes for which multitudes of poloidal harmonics and high radial resolution are required, especially when the kinetic particle responses are fully incorporated. 

Recently, within the GFLDR theoretical framework, a new eigenvalue code was developed to investigate the linear physics of Alfvén eigenmodes in general axisymmetric toroidal geometry\cite{GWeiPoP2024}. With the employment of ballooning-mode representation, it not only relieves the computation resource but also is able to capture the fine radial structure of the mode with minimum effort. The code was initially developed based on ideal magnetohydrodynamic (MHD) equations without EP contribution, so it could only give the real frequency and mode structure of TAE, as well as a small damping rate if the acoustic continuum coupling was considered\cite{GWeiPoP2024}. Adopting the same theoretical framework and similar methodology, in this work, a gyrokinetic version of that code is developed to study the stability of TAE with the energetic particle drive and core plasma Landau damping self-consistently treated in a non-perturbative way. The general particle responses of both circulating and trapped particles are incorporated by means of action-angle approach\cite{FZoncaNJP2015}, and, particularly, the finite Larmor radius (FLR) and finite orbit width effects (FOW) of EPs are fully taken into account. Currently, the code supports Maxwellian, isotropic slowing-down and model anisotropic slowing-down distributions, in order to address the EP physics in a broad range of interest. The possibility of generalization to arbitrary particle distributions is also retained, and will be discussed in our future work. A key important feature of our extended code consists in the systematic implementation of numerical diagnostics, including effective mode structure and resonance structure, thanks to which the code is able to study the physics of wave-particle interaction and power exchange in great detail. In this paper, the ideal MHD approximation, i.e., vanishing parallel electric field, is adopted to simplify the eigenmode equations, which is justified for TAE generally dominated by Alfvénic polarization. As an application of the code, we perform an in-depth study of the triangularity effects on TAE stability based on the reference equilibrium of Divertor Tokamak Test (DTT) facility. It is demonstrated that TAE growth rate can be affected by the triangularity through the modifications of geometric coefficients, resonance condition, as well as the mode frequency and mode structure. Besides, the relative importance of these factors under different circumstances is discussed. The overall effect of negative triangularity for a specific tokamak scenario can be assessed based on these studies. This will be reported in a future publication.

The rest of the paper is organized as follows. In section \ref{Theoretical Framework}, we present the eigenmode equations, the corresponding solution methods, as well as the systematic numerical diagnostics implemented in the code, including the effective mode structure and wave-particle resonance structure. In section \ref{TAE instability in DTT equilibrium}, we briefly introduce the DTT equilibrium and apply the code to analyze the EP driven TAE instability. In section \ref{Effect of triangularity on TAE stability}, we perform an in-depth study of the triangularity effect on TAE stability, where several physical mechanisms are proposed and discussed under different circumstances. Finally, we give a brief summary of the present work and outline the possible future prospects in section \ref{Summary and prospect}. Appendices \ref{AppA} and \ref{AppB} introduce the alternative schemes implemented in the code to solve the linear gyrokinetic equation and the nonlinear eigenvalue problem, respectively.


\section{Theoretical Framework}\label{Theoretical Framework}

\subsection{Eigenmode equations}\label{Eigenmode equations}

For typical gyrokinetic orderings\cite{EFriemanPoF1982,LChenJGR1991,LChenRMP2016}, the plasma fluctuations can generally be described in terms of three scalar fields, i.e., the electrostatic potential $\delta\phi$, the magnetic scalar potential $\delta\psi$, and the parallel magnetic field perturbation $\delta B_\parallel$, where $\delta\psi$ is associated with the parallel vector potential $\delta A_\parallel$ by $c\bm{b}\cdot\nabla\delta\psi=i\omega\delta A_\parallel$. Suppressing the physics of compressional Alfvén waves in low $\beta$ parameter regime of interest, with $\beta$ the ratio between kinetic and magnetic energy densities, $\delta B_\parallel$ can be evaluated by the perturbed perpendicular pressure balance $B_0\delta B_\parallel+4\pi\delta P_\perp=0$\cite{LChenJGR1991}. Building upon the comprehensive theoretical work already established\cite{LChenJGR1991, FZoncaPPCF2006, FZoncaPoP2014a, FZoncaPoP2014b, LChenRMP2016}, we present here the eigenmode equations for investigating EP driven Alfvén instabilities, while omitting the detailed derivations for brevity. Interested readers may refer to references \citenum{FZoncaPoP2014a} and \citenum{FZoncaPoP2014b} for detailed derivations. Assuming the plasma is composed of two components with distinctive energy range: a core or thermal plasma component made of electrons (e) and ions (i), and an energetic particle component (E), the equations governing the general SAW fluctuations then consist of the gyrokinetic vorticity equation 
\begin{equation}\label{VT_eq}
  \begin{aligned} 
    &\left(\partial_{\vartheta}^{2}-\frac{\partial_{\vartheta}^{2}\hat{\kappa}_{\perp}}{\hat{\kappa}_{\perp}}\right)\delta\hat{\Psi}+\frac{{\mathcal J}^{2}B_{0}^{2}}{v_{A}^{2}}\omega(\omega-\omega_{*pi})(\delta\hat{\Phi}_\parallel+\delta\hat{\Psi})\\
    &-8\pi\mathcal{J}^{2}\frac{rB_{0}\partial_rP_0}{q\hat{\kappa}_{\perp}\partial_r\psi}\left(\kappa_{g}\frac{\hat{\bm{\kappa}}_{\perp}\cdot\nabla\psi}{\hat{\kappa}_{\perp}|\nabla\psi|}-\kappa_{n}\frac{rB_{0}}{q\hat{\kappa}_{\perp}|\nabla\psi|}\right)\delta\hat{\Psi}\\
    & -4\pi{\mathcal J}^{2}\frac{rB_{0}}{q\hat{\kappa}_{\perp}\partial_r\psi}\left(\kappa_{g}\frac{\hat{\bm{\kappa}}_{\perp}\cdot\nabla\psi}{\hat{\kappa}_{\perp}|\nabla\psi|}-\kappa_{n}\frac{rB_{0}}{q\hat{\kappa}_{\perp}|\nabla\psi|}\right)\\
    &\times\left\langle m_E\left(\mu B_{0}+v_{\parallel}^{2}\right)\partial_r F_{0E}J_{0E}^{2}\right\rangle _v\delta\hat{\Psi}\\
    &=-\frac{4\pi{\mathcal J}^{2}B_{0}\omega}{k_{\vartheta}c}\left(\kappa_{g}\frac{\hat{\bm{\kappa}}_{\perp}\cdot\nabla\psi}{\hat{\kappa}_{\perp}|\nabla\psi|}-\kappa_{n}\frac{rB_{0}}{q\hat{\kappa}_{\perp}|\nabla\psi|}\right)\\
    &\times\sum_s\left\langle m_s\left(\mu B_{0}+v_{\parallel}^{2}\right)J_{0s}\delta\hat{K}_s\right\rangle _v,
    \end{aligned}
\end{equation}
and the quasi-neutrality condition
\begin{equation}\label{QN_eq}
  \left(1+\frac{T_{0i}}{T_{0e}}\right)\frac{\delta\hat{\Phi}_\parallel}{\hat{\kappa}_\perp}=\frac{T_{0i}}{n_{0i}e_i^2}\sum_{s=e,i}\left\langle e_s\delta\hat{K}_{s}\right\rangle _{v},
\end{equation}
where $\sum_s$ represents summation on all particle species `s', and $\langle\dots\rangle_v$ denotes integration in velocity space. In deriving the above equations, we have made use of the ballooning-mode representation\cite{JConnorPRL1978,RDewarPoF1983} and ignored variations due to the global radial envelope of the fluctuations, with $\vartheta$ denoting the extended poloidal angle and `$\hat{~}$' representing the corresponding quantities in ballooning space. The fluctuating fields have been replaced by $\delta\hat{\Psi}=\hat{\kappa}_\perp\delta\hat{\psi}$ and $\delta\hat{\Phi}_\parallel=\hat{\kappa}_\perp(\delta\hat{\phi}-\delta\hat{\psi})$ for convenience. Moreover, we have adopted the straight magnetic field line coordinates $(\psi,\theta,\zeta)$ with $\psi$ being the poloidal magnetic flux and the Jacobian given by $\mathcal{J}=(\nabla\psi\times\nabla\theta\cdot\nabla\zeta)^{-1}$. Furthermore, the radial-like coordinate $r(\psi)$ has been introduced in equation (\ref{VT_eq}), and the equilibrium magnetic field $\bm{B}_0$ is given by 
\begin{equation}
  \bm{B}_0 = F(\psi)\nabla\phi+\nabla\phi\times\nabla\psi,
\end{equation}
where $\phi$ is the geometric toroidal angle.
The above equations are quite general and retain all the geometric effects. Some of the geometric functions are defined as follows: 
\begin{align*}
  &\bm{\kappa}=\bm{b}_0\cdot\nabla\bm{b}_0,\quad \kappa_{g}=\frac{\bm{\kappa}\cdot\left(\bm{b}_{0}\times\nabla\psi\right)}{|\nabla\psi|},\quad \kappa_{n}=\frac{\bm{\kappa}\cdot\nabla\psi}{|\nabla\psi|}\\
  &k_\vartheta=-nq/r,\quad\hat{\bm{\kappa}}_{\perp}=\frac{\bm{k}_{\perp}}{k_{\vartheta}}=
  s\vartheta\nabla r+r\nabla\theta-\frac{r}{q}\nabla\zeta. 
\end{align*} 
In equation (\ref{VT_eq}), we assume Maxwellian distribution for thermal ions and electrons and ignore their FLR and FOW effects, while the distribution function is kept to be general for EPs, and their FLR and FOW effects are fully retained, consistent with the typical orderings of SAW fluctuations with $|k_\vartheta\rho_i|\ll|k_\vartheta\rho_E|\lesssim 1$, where $\rho_i$ and $\rho_E$ are the drift orbit widths of thermal ions and EPs, respectively. On the left hand side of equation (\ref{VT_eq}), the first three terms represent the field line bending, inertia, and core pressure gradient - curvature coupling, where $\omega_{*pi}=ncT_i/(e\partial_r\psi L_{pi})$ is the thermal ion diamagnetic frequency associated with pressure gradient, with $L_{pi}^{-1}\equiv-\partial_r\ln P_{0i}$ being the scale length of thermal ion pressure, and $P_0=P_{0i}+P_{0e}$. The fourth term represents the EP pressure gradient - curvature coupling term including FLR correction, wherein $F_{0E}$ is the equilibrium particle distribution of EP, and $J_0=J_0(k_\perp\sqrt{2\mu B_0}/\Omega_c)$ is the zero-order Bessel function of the first kind, accounting for the FLR effect, with $\mu=v_\perp^2/(2B_0)$ being the magnetic moment and $\Omega_c=eB_0/(mc)$ the cyclotron frequency. In both the third and fourth terms, we have taken the approximation $\nabla B_0\simeq\bm{\kappa}B_0$, consistent with the well-known cancellation of $\delta B_\parallel$ contribution\cite{LChenJGR1991,FZoncaPoP1999}. The right hand side of equation (\ref{VT_eq}) represents the kinetic plasma compression - magnetic curvature coupling, with the contributions from all particle species included. The contribution of EPs to both the inertia term in equation (\ref{VT_eq}) and the density perturbation in equation (\ref{QN_eq}) have been neglected, due to their much lower density compared to thermal species in fusion plasma with $n_{0E}/n_{0i}\sim\mathcal{O}(10^{-2})$, whereas their contribution to plasma pressure is fully retained noting the typical ordering $P_{0E}/P_{0i}\sim\mathcal{O}(1)$. In equations (\ref{VT_eq}) and (\ref{QN_eq}), the gyrokinetic particle response $\delta\hat{K}$ is obtained by solving the linear gyrokinetic equation 
\begin{equation}\label{GK_eq}
  \begin{aligned}
    &\left(\frac{v_\parallel}{\mathcal{J}B_0}\partial_{\vartheta}-i\omega+i\omega_{d}\right)\delta\hat{K}\\
    &=i\frac{e}{m}QF_{0}\frac{J_0}{\hat{\kappa}_{\perp}}\left(\delta\hat{\Phi}_\parallel+\frac{\omega_{d}}{\omega}\delta\hat{\Psi}\right),
  \end{aligned}
\end{equation}
where $\omega_d=\bm{k}_\perp\cdot\bm{v}_d$ is the magnetic drift frequency, and $\bm{v}_{d}=\Omega^{-1}\bm{b}_0\times\left(\mu\nabla B_{0}+v_{\parallel}^{2}\bm{\kappa}\right)\simeq\Omega^{-1}\bm{b}_0\times\bm{\kappa}\left(\mu B_{0}+v_{\parallel}^{2}\right)$ is the magnetic drift velocity. Furthermore, $QF_0=\omega\partial_{\mathcal{E}}F_0+\Omega^{-1}\bm{b}_0\times\nabla F_0\cdot\bm{k}_\perp$ accounts for the free energy in both velocity and configuration spaces, and ${\mathcal{E}}=v^2/2$ is the energy per unit mass.
Equations (\ref{VT_eq}), (\ref{QN_eq}) and (\ref{GK_eq}) constitute a complete set of equations for investigating the physics of various SAW fluctuations in a broad frequency range. In particular, for the resonant excitation of TAE by EPs considered in this work, further simplification can be made. Due to the high-frequency of TAE with $|\omega|\gg\omega_{ti}\equiv v_{\parallel i}/(\mathcal{J}B_0)$, the quasi-neutrality condition reduces to the ideal MHD approximation with vanishing parallel electric field in the lowest order, i.e., $\delta E_\parallel=0$ or $\delta\hat{\Phi}_\parallel=0$\cite{LChenJGR1991,FZoncaPPCF2006}. Besides, the thermal ion diamagnetic frequency in the inertia term of equation (\ref{VT_eq}) may also be dropped by noting $|\omega_{*pi}/\omega|\ll 1$ for typical plasma parameters. Based on these simplifications, we proceed with the solution of the above equations in the next section.

\subsection{Solution method}\label{Solution method}

In order to solve equation (\ref{GK_eq}) for the perturbed particle distribution, we make use of the action-angle approaches. Following reference \citenum{FZoncaNJP2015}, in axisymmetric toroidal geometry, we introduce three pairs of action angle coordinates, $(m\mu,\alpha)$, $(P_\phi,\phi)$ and $(J,\theta_c)$, where $\alpha$ is the gyrophase, 
\begin{equation}
  P_\phi=\frac{e}{c}\left(F(\psi)\frac{v_\parallel}{\Omega}-\psi\right)
\end{equation}
is the canonical toroidal angular momentum, $J$ and $\theta_c$ are the `second invariant' and the respective conjugate canonical angle defined as 
\begin{equation}\label{J_theta_c}
  J=m\oint v_\parallel dl,\quad \theta_c=\omega_b\int_0^\theta\frac{d\theta'}{\dot{\theta}'},
\end{equation}
where $dl$ is the arc-length along the particle orbit and 
\begin{equation}
  \omega_b=\frac{2\pi}{\oint d\theta/\dot{\theta}}
\end{equation}
is the transit/bounce frequency for circulating/trapped particle. Due to the symmetry of the tokamak geometry, $(\mu,J,P_\phi)$, or equivalently, $(\mathcal{E},\mu,P_\phi)$ for practical convenience, are three constants of motion, and uniquely determine a particle orbit for a given equilibrium magnetic field configuration, together with the sign of $v_\parallel$ which specifies the direction of the circulating particle. Moreover, one has $0<\mu<\mathcal{E}/B_{0,max}$ for circulating particles and $\mathcal{E}/B_{0,max}<\mu<\mathcal{E}/B_{0,min}$ for trapped particles, where $B_{0,min}$ and $B_{0,max}$ are the minimum and maximum values of the magnetic field along the particle trajectory. Given the values of $(\mathcal{E},\mu,P_\phi)$, the guiding center position of a particle can be described as
\begin{equation}\label{orbit}
  \begin{aligned}
    &r=r_c+\tilde{\rho}_c(\theta_c),\\
    &\theta=\sigma\theta_c+\tilde{\Theta}_c(\theta_c),\\
    &\zeta=\bar{\omega}_d\tau+\sigma\bar{q}\theta_c+\tilde{\Xi}_c(\theta_c),
  \end{aligned}
\end{equation}
where $r_c=r(\bar{\psi})$, $\bar{q}=q(\bar{\psi})$, and 
\begin{equation}
  \bar{\psi}=\frac{\omega_b}{2\pi}\oint\psi\frac{d\theta}{\dot{\theta}}
\end{equation}
denotes the orbit averaged magnetic flux. $\bar{\omega}_d$ is the toroidal precession frequency defined as 
\begin{equation}
  \bar{\omega}_d=\frac{\omega_b}{2\pi}\oint\left(\dot{\zeta}-\bar{q}\dot{\theta}\right)\frac{d\theta}{\dot{\theta}},
\end{equation}
and $\tau=\theta_c/\omega_b$ is a time-like variable. Besides, $\tilde{\rho}_c$, $\tilde{\Theta}_c$ and $\tilde{\Xi}_c$ are $2\pi$ periodic functions of $\theta_c$ defined as 
\begin{equation}\label{periodic_functions}
  \begin{aligned}
    &\tilde{\rho}_c(\theta_c)=\left(\frac{Fv_{\parallel}}{\Omega}-\frac{\omega_b}{2\pi}\oint\frac{Fv_{\parallel}}{\Omega}\frac{d\theta}{\dot{\theta}}\right)\frac{1}{d\psi/dr}\Big|_{r_c},\\
    &\tilde{\Theta}_c(\theta_c)=\theta-\sigma\theta_c,\\
    &\tilde{\Xi}_c(\theta_c)=\int_0^{\theta}\left(\frac{\dot{\zeta}}{\dot{\theta}}-\bar{q}\right)d\theta-\bar{\omega}_{d}\tau+\bar{q}(\theta-\sigma\theta_{c}),
  \end{aligned}
\end{equation}
with $\sigma=v_\parallel/|v_\parallel|$ for circulating particles, and $\sigma=0$ for trapped particles. All these characteristic frequencies and periodic functions in equation (\ref{orbit}) can be obtained by solving the guiding center equation of motion\cite{ABrizardRMP2007, JcaryRMP2009} 
\begin{equation}\label{motion_eq}
  \dot{\bm{X}}=\frac{B_{0}}{B_{\parallel}^{*}}\left(v_{\parallel}\bm{b}_0+\frac{v_{\parallel}}{\Omega}\nabla\times(v_{\parallel}\bm{b}_0)\right),
\end{equation}
with 
\begin{equation}
  B_{\parallel}^{*}=B_{0}+\frac{mc}{e}\bm{b}_0\cdot(\nabla\times\bm{b}_0)v_{\parallel}\simeq B_0.
\end{equation}
For our local eigenmode analysis corresponding to fixed $r_c$, we solve equation (\ref{motion_eq}) for the particle orbit with the magnetic field evaluated at the reference flux surface $\bar{\psi}$, so the radial variation of equilibrium magnetic field on the scale of particle drift orbit width is ignored, consistent with the ordering $\rho_E/L_B\ll 1$ in fusion plasmas, where $L_B\sim R_0$ denotes the scale length of equilibrium magnetic field nonuniformity. Nevertheless, the particle drift motions are fully taken into account through the second term in the bracket of equation (\ref{motion_eq}), and play a crucial role in determining mode stability. More accurate equilibrium particle orbit calculation can be readily implemented in our numerical scheme at the expense of more intensive use of computational resources. This has been verified to actually yield an $\mathcal{O}(\rho_E/L_B)$ correction and, thus, is consistently neglected in the present analysis.

With the coefficients and functions in equation (\ref{orbit}) parameterized by the constants of motion, the linear gyrokinetic equation can be solved most conveniently by taking the drift/banana center transformation $\delta\hat{K}=e^{-i\hat{Q}_B}\delta\hat{K}_B$, and yields 
\begin{equation}\label{GK_eq_drift_center}
  \left(\omega_{b}\partial_{\theta_{c}}-i\omega+in\bar{\omega}_{d}\right)\delta\hat{K}_{B}=i\frac{e}{m}e^{i\hat{Q}_{B}}QF_{0}J_{0}\frac{\omega_{d}}{\omega}\frac{\delta\hat{\Psi}}{\hat{\kappa}_{\perp}},
\end{equation}
where we have dropped the contribution of $\delta\hat{\Phi}_\parallel$, and the shift operator $\hat{Q}_B$ is defined as 
\begin{equation}
  \hat{Q}_{B}=\tilde{\rho}_{c}(\theta_{c})k_{\vartheta}s\vartheta+n\tilde{\Xi}_{c}(\theta_{c})-n\bar{q}\tilde{\Theta}_{c}(\theta_{c}),
\end{equation}
which essentially accounts for the FOW effect. In equation (\ref{GK_eq_drift_center}), the left hand side is a linear operator with constant coefficients, and the right hand side is the fluctuating field shifted to the magnetic drift/banana center coordinates, which can be considered as the effective mode structure that is actually experienced by the particle along its orbit. The sign of $v_\parallel$ is implicitly embedded in the mapping relation between $\theta_c$ and $\vartheta$ (or $\theta$). For circulating particles, by directly applying the Fourier and inverse Fourier transformations on equation (\ref{GK_eq_drift_center}) in $\theta_c$ space, we obtain 
\begin{equation}\label{solution_circulating}
  \begin{aligned}
    \delta\hat{K}_{B}(\theta_{c})=&\frac{e}{m}\frac{QF_{0}}{\omega}\int_{-\infty}^{\infty}dk\frac{e^{ik\theta_{c}}}{k\omega_{b}-\omega+n\bar{\omega}_{d}}\\
    \times&\int_{-\infty}^{\infty}\frac{d\theta_{c}'}{2\pi}e^{i\hat{Q}_{B}'}\frac{J_{0}'\omega_d'}{\hat{\kappa}_{\perp}'}\delta\hat{\Psi}'e^{-ik\theta_{c}'},
  \end{aligned}
\end{equation}
where the definition of $\theta_c(\vartheta)$ has been extended to $(-\infty, \infty)$. 
For trapped particles, the two bounce angles $\theta_1$ and $\theta_2$ can be introduced by the condition $1-\mu B_0(\theta_{1,2})/\mathcal{E}=0$, and the closed bounce orbit in $\vartheta$ space, i.e., $\theta_1\rightarrow\theta_2\rightarrow\theta_1$, can be mapped into $\theta_c$ space as $-\pi/2\rightarrow\pi/2\rightarrow 3\pi/2$ according to equation (\ref{J_theta_c}). This mapping relation ensures that the particle response naturally satisfies the periodic boundary condition over the $2\pi$ interval $-\pi/2\leq\theta_c\leq 3\pi/2$. Therefore, $\delta\hat{K}_B$ can be decomposed into Fourier series $\delta\hat{K}_B=\sum_{k}\delta\hat{K}_{B,k}e^{ik\theta_c}$, with $k$ taking integer values, and we obtain, from equation (\ref{GK_eq_drift_center}), 
\begin{equation}\label{solution_trapped}
  \begin{aligned}
    \delta\hat{K}_{B}(\theta_{c})=&\frac{e}{m}\frac{QF_{0}}{\omega}\sum_{k\in\mathbb{Z}}\frac{e^{ik\theta_{c}}}{k\omega_{b}-\omega+n\bar{\omega}_{d}}\\
    \times&\int_{-\pi/2}^{3\pi/2}\frac{d\theta_{c}'}{2\pi}e^{i\hat{Q}_{B}'}\frac{J_{0}'\omega_d'}{\hat{\kappa}_{\perp}'}\delta\hat{\Psi}'e^{-ik\theta_{c}'},
  \end{aligned}
\end{equation}
where the integration in $\theta_c$ corresponds to the orbit average along the closed banana orbit in the poloidal plane. The solutions in other intervals can be obtained by shifting $\theta_c$ (and correspondingly $\vartheta$) by $2\pi$ in equation (\ref{solution_trapped}). It should be noted that the periodic boundary condition that we imposed in $\theta_c$ space is equivalent to the conventional boundary condition in $\vartheta$ space: $\delta\hat{K}^+(\vartheta=\theta_1)=\delta\hat{K}^-(\vartheta=\theta_1)$ and $\delta\hat{K}^+(\vartheta=\theta_2)=\delta\hat{K}^-(\vartheta=\theta_2)$\cite{WTangNF1980,GRewoldtPoFB1982,LChenJGR1991,GFuPoFB1992,GFuPoFB1993}, where the superscript $\pm$ denotes the sign of $v_\parallel$. 

With the solution of $\delta\hat{K}_B$, the guiding center response $\delta\hat{K}$ can be directly obtained through the pull back operator $e^{-i\hat{Q}_B}$. Compared with the conventional method that directly integrates equation (\ref{GK_eq}) or equation (\ref{GK_eq_drift_center}) along the unperturbed orbits\cite{WTangNF1980,GRewoldtPoFB1982,GFuPoFB1992,GFuPoFB1993,YLiPoP2020} (see Appendix \ref{AppA}), the Fourier spectrum methods we proposed here leverages the periodicity of the equilibrium orbits in $\theta_c$ space described by equation (\ref{orbit}), and enables efficient computation of particle responses via fast Fourier transformation (FFT) algorithm. Moreover, equations (\ref{solution_circulating}) and (\ref{solution_trapped}) can readily reproduce the resonance condition 
\begin{equation}\label{resonance_condition}
\omega=n\bar{\omega}_d+k\omega_b,
\end{equation}
where the value of $k$ is related to the Fourier spectrum of the effective mode structure. In particular, $k$ is integer for trapped particles due to the periodicity of particle response that we mentioned earlier. 

Substituting the solutions of $\delta\hat{K}$ for both circulating and trapped particles back into the vorticity equation and carrying out the velocity space integration, we obtain the corresponding eigenvalue problem. This problem is intrinsically a nonlinear eigenvalue problem and is hard to solve directly. However, for TAE instabilities excited by EPs, the contribution of the kinetic compressibility is usually much smaller than the fluid potential\cite{GFuPoFB1992,GFuPoFB1993}, inspired by which the following iteration procedure is proposed. For the sake of brevity, we formally rewrite the vorticity equation as 
\begin{equation}\label{VT_eq_simplified}
  \left[\partial_\vartheta^2+V_f(\omega;\vartheta)\right]\delta\hat{\Psi}=\text{KC}(\omega,\delta\hat{\Psi};\vartheta),
\end{equation}
where $V_f(\omega;\vartheta)$ represents the fluid-like potential on the left hand side of the vorticity equation, and $\text{KC}(\omega,\delta\hat{\Psi};\vartheta)$ is the kinetic compression term. First, we solve the vorticity equation in the fluid limit by neglecting the kinetic compression term. The obtained mode frequency and mode structure, denoted as $\omega_0$ and $\delta\hat{\Psi}_0$, are then substituted back into the kinetic compression term to update the eigenmode equation
\begin{equation}
  \left[\partial_\vartheta^2+V_f(\omega_1;\vartheta)\right]\delta\hat{\Psi}_1=\text{KC}(\omega_0,\delta\hat{\Psi}_0;\vartheta),
\end{equation}
where $\omega_1$ and $\delta\hat{\Psi}_1$ are the updated mode frequency and mode structure. We iterate this procedure until the solutions of eigenvalue and eigenfunction converge, and during each iteration, the equation can be easily solved by the shooting method. 
The above iteration approach is particularly efficient for studying the EP driven TAE instabilities, and it is even not restricted to the condition that kinetic effects are perturbative, as long as the iteration converges. 
However, because of the inconsistent treatment of mode frequency and mode structure on the left and right hand sides of equation (\ref{VT_eq_simplified}), this approach may sometimes encounter convergence difficulties when the kinetic effects become sufficiently large and significantly alter the mode frequency and mode structure. In order to maintain the general capability of our code, we have also implemented the algorithm based on finite element method, as introduced in Appendix \ref{AppB}.

Consistent with former discussions, fast estimation of TAE frequency shift and growth rate induced by kinetic effects can be achieved by the expansion of equation (\ref{VT_eq_simplified}) around the solution in fluid limit\cite{GFuPoFB1992}. The complex frequency shift $\Delta\omega$ introduced by the kinetic compression term is then given by 
\begin{equation}\label{perturbative_eq}
  \frac{\Delta\omega}{\omega_{0}}=\frac{\delta W_{k}}{2\int_{-\infty}^{\infty}d\vartheta\frac{\omega_{0}^{2}{\cal J}^{2}B_{0}^{2}}{v_{A}^{2}}\left|\delta\hat{\Psi}_{0}\right|^{2}},
\end{equation}
where $\delta W_{k}\equiv\int_{-\infty}^{\infty}d\vartheta\delta\hat{\Psi}_{0}^{*}\text{KC}(\omega_{0},\delta\hat{\Psi}_{0};\vartheta)$ \cite{LChenRMP2016,FZoncaPoP1996,FZoncaPRL1992} denotes the generalized potential energy contributed by the kinetic compressibilities of all particle species, with $\omega_0$ and $\delta\hat{\Psi}_0$ being the TAE solutions in fluid limit. Apparently, the above equation is valid only when the contribution of particle kinetic compressibilities is perturbative, i.e., $|\Delta\omega|\ll|\Delta\omega_{SAW}|$\cite{FZoncaNJP2015,TWangPoP2018}, where $\Delta\omega_{SAW}$ is the frequency difference between TAE and neighboring mode including the SAW continuous spectrum. Although this assumption does not always hold even for TAE, the perturbative analysis could be useful in many circumstances of interest.

\subsection{Physics based numerical diagnostics}\label{Physics based numerical diagnostics}

Here, we introduce the numerical diagnostics that have been developed to elucidate the fundamental physics of the interaction between TAE and different plasma components. All the numerical calculations in this section, for illustration purpose, are performed based on the DTT equilibrium that will be introduced in section \ref{TAE instability in DTT equilibrium} with EPs satisfying a model isotropic slowing down distribution, and TAE toroidal mode number is taken to be $n=12$. While these choices may seem specific to the DTT case that will be studied later, both magnetic geometry and plasma profiles as well as mode number are paradigmatic for reactor relevant burning plasma scenarios.

\subsubsection{Effective mode structure}

\begin{figure}[]
  \centering
  \includegraphics[scale=0.6]{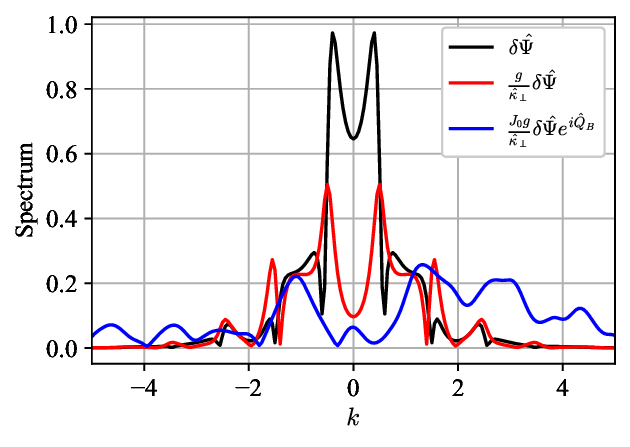}
  \caption{$k$ spectra of $\delta\hat{\Psi}$, $g\delta\hat{\Psi}/\hat{\kappa}_\perp$, $e^{i\hat{Q}_B}J_0g\delta\hat{\Psi}/\hat{\kappa}_\perp$ for circulating particle in the whole ballooning space.}\label{spectrum_pass} 
\end{figure}

\begin{figure}[]
  \centering
  \includegraphics[scale=0.6]{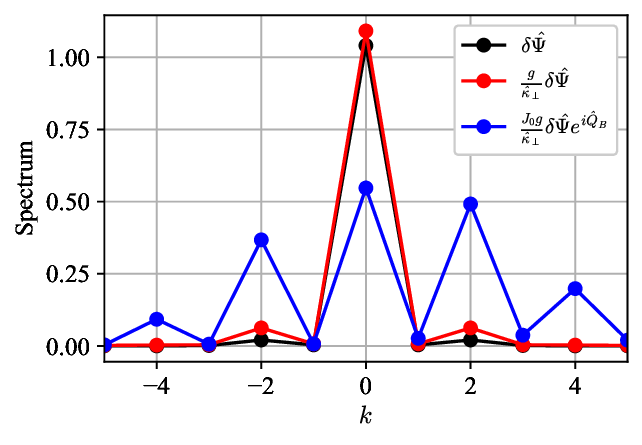}
  \caption{$k$ spectra of $\delta\hat{\Psi}$, $g\delta\hat{\Psi}/\hat{\kappa}_\perp$, $e^{i\hat{Q}_B}J_0g\delta\hat{\Psi}/\hat{\kappa}_\perp$ for trapped particle in $[-\theta_b,\theta_b]$ interval.}\label{spectrum_trap} 
\end{figure}

One of the key numerical diagnostics in the code is the representation of effective mode structure and the corresponding spectrum. As discussed in section \ref{Solution method}, the wave-particle interaction is determined by both the resonance condition, equation (\ref{resonance_condition}), and the effective mode structure, which can be defined as $e^{i\hat{Q}_B}J_0g\delta\hat{\Psi}/\hat{\kappa}_\perp$ according to equations (\ref{solution_circulating}) and (\ref{solution_trapped}). Here, 
\begin{equation}\label{g_func}
g=\frac{R_{0}\bar{B}_0}{B_0}\bm{b}_{0}\times\bm{\kappa}\cdot\hat{\bm{\kappa}}_{\perp}
\end{equation}
accounts for the explicit $\vartheta$-dependence in magnetic curvature drift, so that $\omega_d=k_\vartheta(R_0\bar{\Omega})^{-1}(\mu B_0+v_\parallel^2)g$ with $\bar{\Omega}=e\bar{B}_0/(mc)$. Figures \ref{spectrum_pass} and \ref{spectrum_trap} show the Fourier spectra of the effective mode structures with and without the effect of FLR and FOW for circulating and trapped particles, respectively. For the purpose of demonstration, we use TAE mode structure $\delta\hat{\Psi}$ obtained from equation (\ref{VT_eq}) in fluid limit, which is shown by the blue dashed line in figure \ref{mode_structure}. In addition, the particle orbits are calculated with $(\mathcal{E},\lambda)=(\mathcal{E}_0,0.1)$ for the circulating particle in figure \ref{spectrum_pass}, and $(\mathcal{E},\lambda)=(\mathcal{E}_0,1.1)$ for the trapped particle in figure \ref{spectrum_trap}, where $\mathcal{E}$ is the EP birth energy and $\lambda\equiv\mu\bar{B}_0/\mathcal{E}$ is the pitch angle. For the circulating particle, the spectrum of $\delta\hat{\Psi}$ peaks around $k=\pm1/2$, consistent with the well-known fact that TAE is localized around $nq-m=\pm1/2$. The coupling with magnetic curvature generates higher $k$ components around $\pm3/2$ and $\pm5/2$, as shown by the red line. For thermal ions and electrons with orbit widths much smaller than typical TAE wavelength, $g\delta\hat{\Psi}/\hat{\kappa}_\perp$ essentially plays the role as effective mode structure experienced by the particle, and is mostly identical to the mode structure in laboratory frame except for the correction due to the differences between $\vartheta$ (or $\theta$) and $\theta_c$. However, for EPs with relatively large orbit widths, the corrections of FLR and FOW effects, accounted for by $J_0e^{i\hat{Q}_B}$, must be taken into account properly. As a consequence, the spectrum of effective mode structure is greatly distorted, as shown by the blue line in figure \ref{spectrum_pass}, which clearly demonstrates that the incorporation of FLR and FOW effects reduces the spectrum amplitude (weakening the wave-particle coupling strength) while simultaneously broadens the spectrum range (involving more particles into resonance)\cite{GFuPoFB1992}. As a consequence, the overall effects of FLR and FOW on TAE stability depend on the specific particle distribution in velocity space. 

For the trapped particle, the effective mode structure is defined inside the bounce interval $[\theta_1, \theta_2]$, and thus is sensitive to the pitch angle of the particle. In figure \ref{spectrum_trap}, we choose $\lambda=1.1$, which is close to the deeply trapped particle region. The spectrum of $\delta\hat{\Psi}$ is dominated by the bounce-averaged $k=0$ component, and the coupling with magnetic curvature has little impact on the spectrum distribution, consistent with deeply trapped particle region that has been chosen. However, for particle with large orbit width, the introduction of FLR and FOW effects significantly modifies the spectrum, where the dominant $k=0$ component is reduced while the $k=\pm2$ harmonics are greatly enhanced. From these results, it can be speculated that trapped particle responses are dominant by precession resonance, as well as the low order bounce resonances in the presence of FLR and FOW effects, while high order bounce resonances are less efficient due to the decay of the Fourier harmonics. It is noteworthy that the amplitudes of odd harmonics are significantly smaller than the even ones, consistent with the fact that the fields experienced by a trapped particle in the inner and outer legs of its banana orbit are nearly identical in an up-down symmetric equilibrium. Furthermore, the broken symmetry under parity transformation of the blue line with respect to $k$ in figure \ref{spectrum_trap} (also in figure \ref{spectrum_pass}) arises from the phase variation introduced by the FOW effect, i.e., $e^{i\hat{Q}_B}$. 

\subsubsection{Wave-particle resonance structure}

Another numerical diagnostics of crucial importance in our code is the wave-particle resonance structure in velocity space. Together with the effective mode structures, it enables us to investigate wave-particle interaction in great detail. Suggested by equation (\ref{perturbative_eq}), the value of $\delta W_k$ contains the information of frequency shift and growth rate induced by the kinetic effects\cite{LChenRMP2016,FZoncaPoP1996,FZoncaPRL1992}. In particular, the imaginary part of $\delta W_k$ is proportional to the energy exchange between particles and wave through the resonant interaction. Notice that $\delta W_k$ is essentially an integral with respect to $\mathcal{E}$, $\mu$ and $\vartheta$. By plotting the integrand of $\delta W_k$ in $(\mathcal{E},\mu)$ space, we will be able to show the resonance properties of different particle species and identify which type of equilibrium particle orbits predominantly contribute to the drive or damping of the mode. Due to the fundamentally distinct behaviors of circulating and trapped particles, as well as the differences among particle species, we will study their resonance structures separately in the rest of the section. For the present equilibrium at $\rho_{tor} = 0.54$, particles with $0<\lambda<0.81$ are circulating, while those with $0.81<\lambda<1.17$ are trapped. 

\begin{figure}[htbp]
    \centering
    \begin{subfigure}[b]{0.48\textwidth}
        \includegraphics[width=\textwidth]{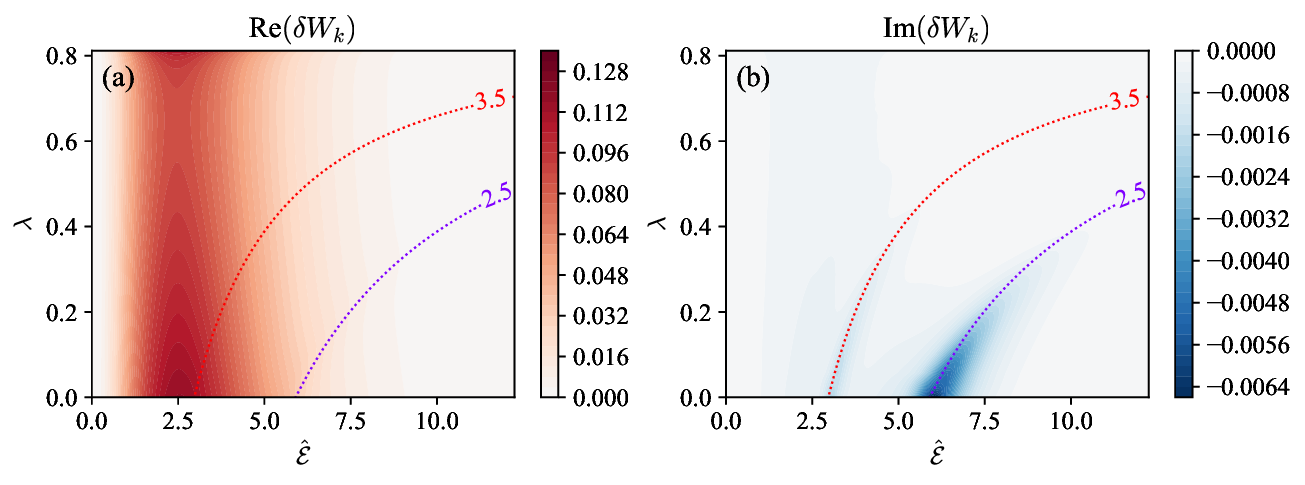}
    \end{subfigure}
    \hspace{0\textwidth}
    \begin{subfigure}[b]{0.48\textwidth}
        \includegraphics[width=\textwidth]{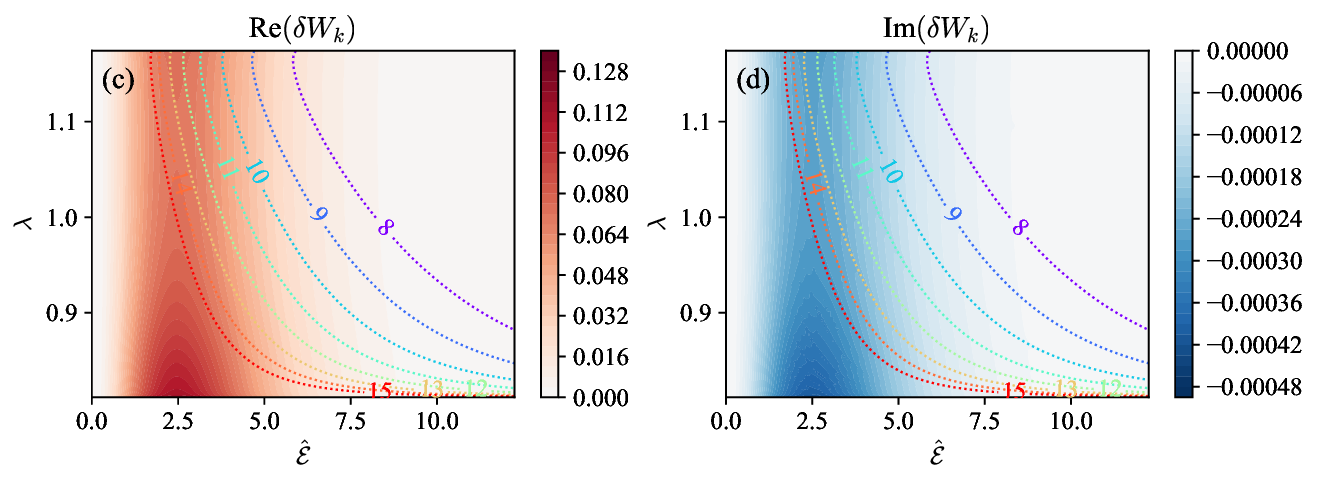}
    \end{subfigure}
    \caption{Structures of the integrand of $\delta W_k$ in $(\hat{\mathcal{E}},\lambda)$ space for (a, b) circulating and (c, d) trapped thermal ions. The colored dotted lines represent the contours of $k$ satisfying the resonance condition $\omega=n\bar{\omega}_d+k\omega_b$.}
    \label{deltaWk_Ti}
\end{figure}

\emph{Thermal ion responses.} First of all, figures \ref{deltaWk_Ti}(a) and (b) show the integrand of $\delta W_k$ due to circulating thermal ions in the space of normalized energy $\hat{\mathcal{E}}\equiv2m\mathcal{E}/T_i$ and pitch angle $\lambda$. The colored dotted lines represent the contours of $k$ satisfying the resonance condition $\omega=n\bar{\omega}_d+k\omega_b$, where both $\bar{\omega}_d$ and $\omega_b$ are the functions of $\hat{\mathcal{E}}$ and $\lambda$. For circulating thermal ions, the real part of $\delta W_k$ is positive and dominated by the non-resonant response, while the imaginary part is negative and dominated by the resonant response, and its amplitude is very small, in consistency with the $v_{ti}/v_A\ll1$ ordering. According to equation (\ref{perturbative_eq}), these results imply that the kinetic compressibility of circulating thermal ions primarily gives rise to a positive frequency shift and a small damping rate to TAE. Furthermore, the maximum value position of $\text{Re}(\delta W_k)$ is located around $\lambda=0$, suggesting that well circulating approximation along with fluid expansion may be a good approximation when solving the kinetic response of circulating thermal ions. The dominant contribution to the $\text{Im}(\delta W_k)$ comes from the $k=5/2$ resonance, which, consistent with the spectrum of $g\delta\hat{\Psi}/\hat{\kappa}_\perp$ in figure \ref{spectrum_pass}, has a very small amplitude. Lower order resonances $k=1/2,3/2$ are suppressed exponentially following the scaling law $\propto \exp(-v_{res}^2/v_t^2)$, where $v_{res}\simeq v_A/(2k)$ is the ion velocity to resonate with TAE with $\omega\simeq v_A/(2qR_0)$. For trapped thermal ions, the integrand of $\delta W_k$ exhibits similar pattern with that of circulating particles, as shown in figures \ref{deltaWk_Ti}(c) and (d), where the non-resonant particle responses dominate. The resonance lines shows that only high order bounce resonances can occur because of the small $n\bar{\omega}_d$ and $\omega_b$ of thermal ions, while such high order bounce resonances are very inefficient due to the rapid decay of the high $k$ components in the spectrum of $g\delta\hat{\Psi}/\hat{\kappa}_\perp$. 

\begin{figure}[htbp]
    \centering
    \begin{subfigure}[b]{0.48\textwidth}
        \includegraphics[width=\textwidth]{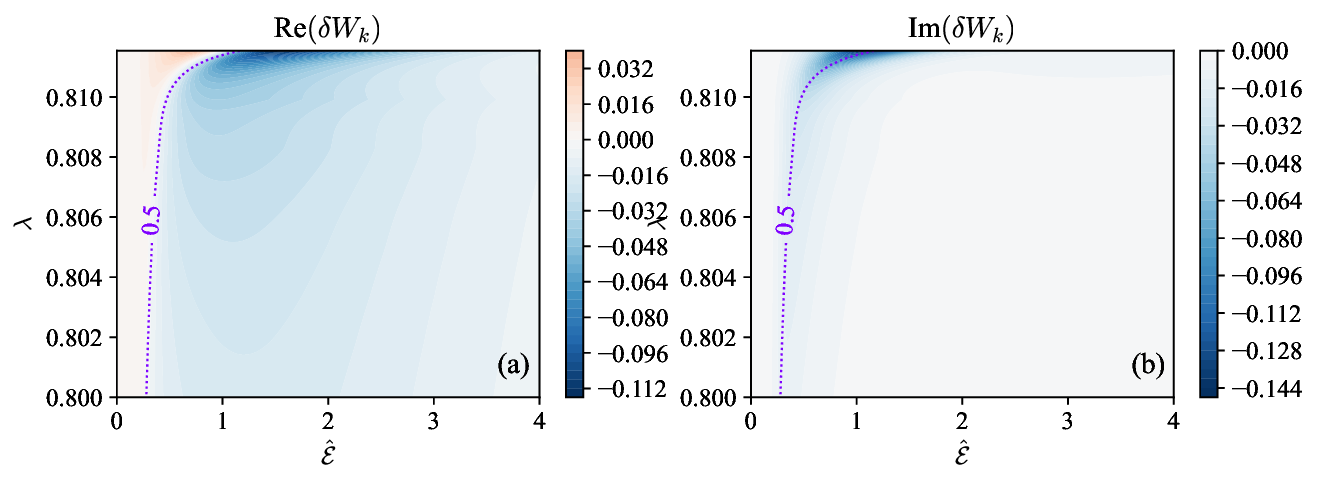}
    \end{subfigure}
    \hspace{0\textwidth}
    \begin{subfigure}[b]{0.48\textwidth}
        \includegraphics[width=\textwidth]{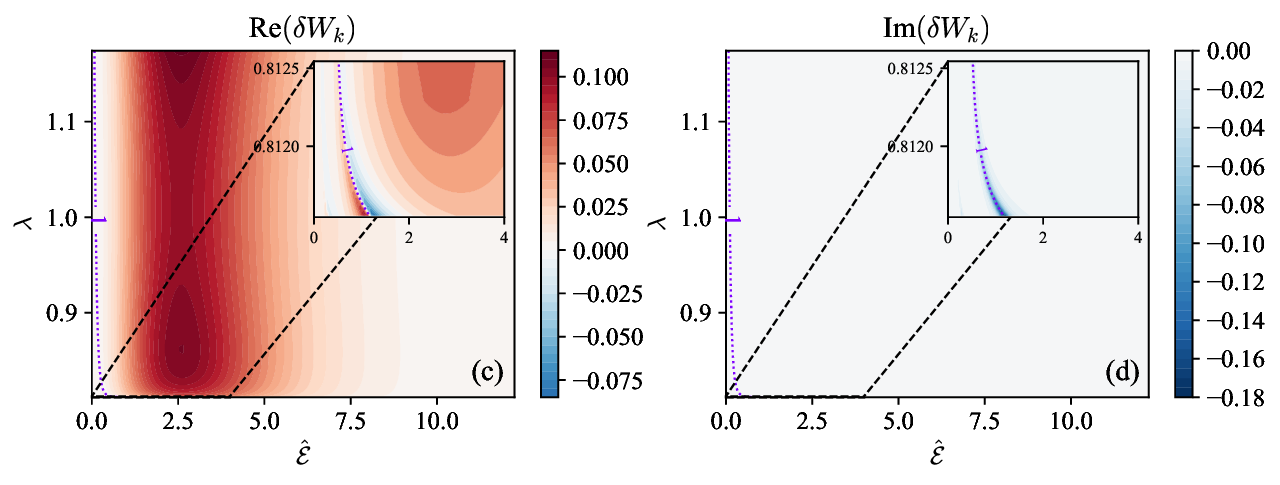}
    \end{subfigure}
    \caption{Structures of the integrand of $\delta W_k$ in $(\hat{\mathcal{E}},\lambda)$ space for (a, b) circulating and (c, d) trapped thermal electrons. The colored dotted lines represent the contours of $k$ satisfying the resonance condition $\omega=n\bar{\omega}_d+k\omega_b$.}
    \label{deltaWk_Te}
\end{figure}

\emph{Thermal electron responses.} For circulating thermal electrons, as illustrated in figures \ref{deltaWk_Te}(a) and (b), the amplitudes of both real and imaginary of $\delta W_k$ are small, and are limited in a narrow region in the velocity space near the boundary between circulating and trapped particles (note the range of the vertical axis), which means the particle responses are nearly adiabatic due to their high speed along the magnetic field $v_{\parallel e}\gg v_A$. Figures \ref{deltaWk_Te}(a) and (b) indicates that the dominant contribution comes from the $k=1/2$ resonance, while higher order resonances corresponding to lower energy are suppressed algebraically due to the $\mathcal{E}^{5/2}$ dependence in the integrand. Similar pattern is observed for the resonant trapped thermal electrons in figure \ref{deltaWk_Te}(c) and (d), with a tiny $k=1$ resonance localized in a very narrow region. The difference is that the non-resonant trapped particles has a relatively large contribution to $\text{Re}(\delta W_k)$ because of their smaller bounce frequencies. Therefore, the trapped electrons can induce a finite frequency shift to the mode. Although the wave-particle resonances are weak for both thermal ions and electrons, the underlying reasons are opposite: thermal ions are too slow, while thermal electrons are too fast, with respective to the Alfvén speed. Thus, it can be concluded that thermal ion Landau damping becomes more important in high-$\beta$ plasma\cite{RBettiPoFB1991,RBettiPoFB1992}, while thermal electron Landau damping becomes more important in low-$\beta$ plasma\cite{GFuPoFB1989}. 

\begin{figure}[htbp]
    \centering
    \begin{subfigure}[b]{0.48\textwidth}
        \includegraphics[width=\textwidth]{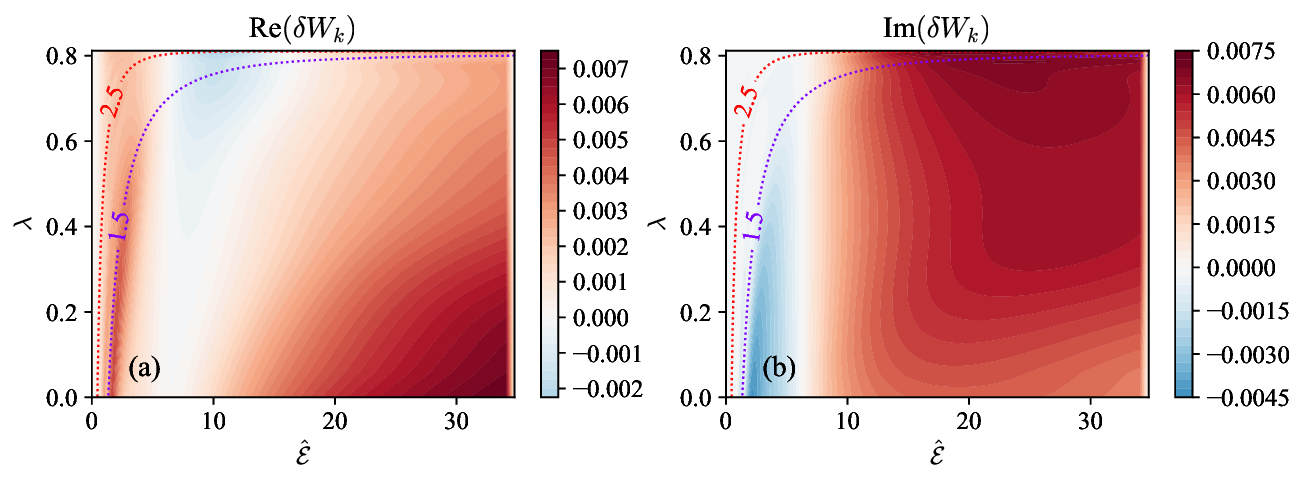}
    \end{subfigure}
    \hspace{0\textwidth}
    \begin{subfigure}[b]{0.48\textwidth}
        \includegraphics[width=\textwidth]{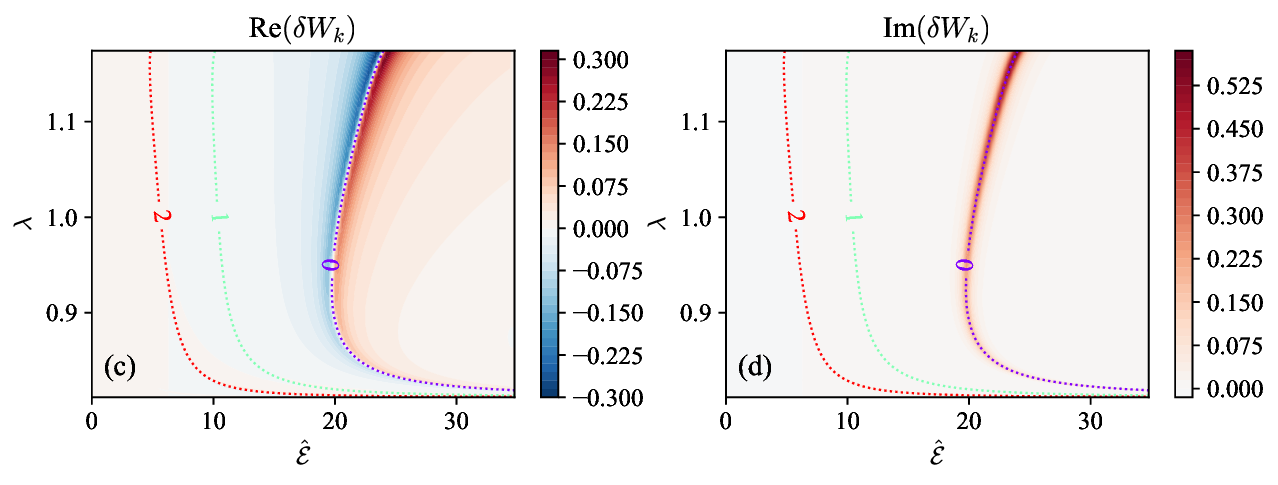}
    \end{subfigure}
    \caption{Structures of the integrand of $\delta W_k$ in $(\hat{\mathcal{E}},\lambda)$ space for (a, b) circulating and (c, d) trapped EPs. The colored dotted lines represent the contour lines of $k$ satisfying the resonance condition $\omega=n\bar{\omega}_d+k\omega_b$.}
    \label{deltaWk_EP}
\end{figure}

\emph{EP responses.} Figures \ref{deltaWk_EP}(a) and (b) show the resonance structure of circulating EPs in $(\hat{\mathcal{E}},\lambda)$ space, where the definition of $\hat{\mathcal{E}}$ is modified to $\hat{\mathcal{E}}\equiv\mathcal{E}/\mathcal{E}_c$, with $\mathcal{E}_c$ being the critical energy. Since the integrand of $\delta W_k$, proportional to $\mathcal{E}^{5/2}/(\mathcal{E}^{3/2}+\mathcal{E}_c^{3/2})$, monotonically increases with $\mathcal{E}$, it is observed that most of the EP contributions come from the high energy region in the velocity space, different from the case of Maxwellian distribution. Besides, figure \ref{deltaWk_EP}(b) exhibits no peak structures along the half-integer resonance lines as a consequence of the large orbit widths of EP, and it is also consistent with the Fourier spectrum of $e^{i\hat{Q}_B}J_0g\delta\hat{\Psi}/\hat{\kappa}_\perp$ in figure \ref{spectrum_pass}. Whereas for trapped EPs, the resonance structures depicted in figure \ref{deltaWk_EP}(c) and (d) are clearly visible, especially the $k=0$ precession resonance. According to figure \ref{spectrum_trap}, the $k=2$ bounce resonance also exists, but has much smaller amplitude due to the algebraic decay of the integrand with respect to energy. From the results of figures \ref{deltaWk_Ti}, \ref{deltaWk_Te} and \ref{deltaWk_EP}, combined with equation (\ref{perturbative_eq}), it follows the conclusion that EPs contribute little to the real frequency, but substantially to the growth rate of TAE compared with thermal ions and electrons.

\section{TAE instability in DTT equilibrium}\label{TAE instability in DTT equilibrium}

Consistent with the analysis in the previous section, we consider a realistic DTT equilibrium to illustrate the spectral features of TAE in actual conditions, but assume a simple model distribution of EP represented by a simple isotropic slowing down injected at 3.52 MeV, since for our scopes the effect of EP is merely to provide a finite drive. The capability of our code to explore stability with realistic distribution functions provided by numerical computations will be reported in future work.

\begin{figure}[htbp]
    \centering
    \begin{subfigure}[]{0.205\textwidth}
        \includegraphics[width=\textwidth]{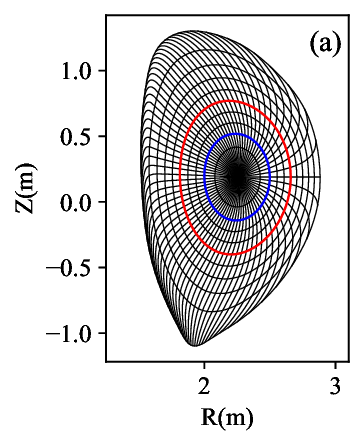}
    \end{subfigure}
    \begin{subfigure}[]{0.27\textwidth}
        \includegraphics[width=\textwidth]{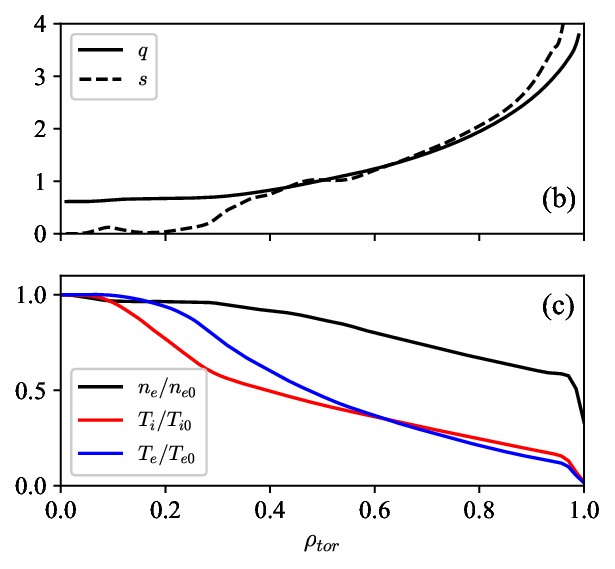}
    \end{subfigure}
    \caption{(a) Cross section of DTT reference equilibrium, where the red and blue lines represent the radial positions of $\rho_{tor}=0.55$ and $\rho_{tor}=0.32$, respectively. (b) Radial profiles of safety factor $q$ and magnetic shear $s$. (c) Radial profiles of electron density, electron and ion temperatures normalized to their values on axis. The horizontal axis $\rho_{tor}$ is the normalized radius related to toroidal flux.}\label{equilibrium} 
\end{figure}

Here, we briefly introduce the equilibrium parameters for the numerical application of the newly developed code. The equilibrium we adopt is one of the reference equilibria of DTT, which is originally constructed using the free boundary equilibrium evolution code CREATE-NL26\cite{RAlbaneseFED2003} and further refined using the high-resolution equilibrium solver CHEASE\cite{HLutjensCPC1996}. The equilibrium is post processed by FALCON code\cite{MFalessiPoP2019_continuum,MFalessiJPP2020}, which constructs Boozer coordinates and generates the corresponding metric tensor required for eigenmode calculation. 
In this equilibrium, the on-axis thermal electron density is $2.39\times 10^{20}\text{m}^{-3}$, and the on-axis thermal electron and ion temperatures are $13.04\text{KeV}$ and $9.92\text{KeV}$, respectively. The cross section and thermal plasma radial profiles are depicted in figure \ref{equilibrium}, where we have introduced the normalized radius $\rho_{tor}$ as flux coordinate. For the study of EP driven TAE instability in this work, we assume an isotropic slowing-down distribution for EPs in the form of 
\begin{equation}
  \frac{f_{0E}}{n_{0E}}\propto\frac{\Theta({\cal E}-{\cal E}_{0})}{{\cal E}^{3/2}+{\cal E}_{c}^{3/2}},
\end{equation}
where $\Theta$ denotes the Heaviside step function. The birth energy $\mathcal{E}_0$ is taken to be the fusion energy of alpha particles, i.e., 3.52 MeV, and critical energy $\mathcal{E}_c$ is given by Stix expression\cite{TStixPoP1972}. $n_{0E}$ is the equilibrium density of EPs, and for our local eigenmode analysis in this work, we take $n_{0E}=1.0\times 10^{18}\text{m}^{-3}$ and $R_0\partial_{\rho_{tor}}\ln n_{0E}=5.0\text{m}$. 
We note that, while DTT will not operate with deuterium and tritium, and there won't be fusion alpha particles, the DTT equilibrium and alpha particles as EP source are adopted here simply as a reference for our numerical analysis, while the underlying physics we would like to elucidate is rather general and not limited to this specific scenario. As anticipated above, our aim is to illuminate the properties of TAE spectra as paradigm of AE in reactor relevant plasmas, capturing how magnetic geometry is interlinked with particle orbits, wave-particle resonance conditions and plasma kinetic compressibility. The choice of fusion alpha particles as an isotropic slowing down source, in particular, allows us to enhance the FLR/FOW effect of EP, which was discussed in the previous section.

\begin{figure}[htbp]
  \includegraphics[width=0.40\textwidth]{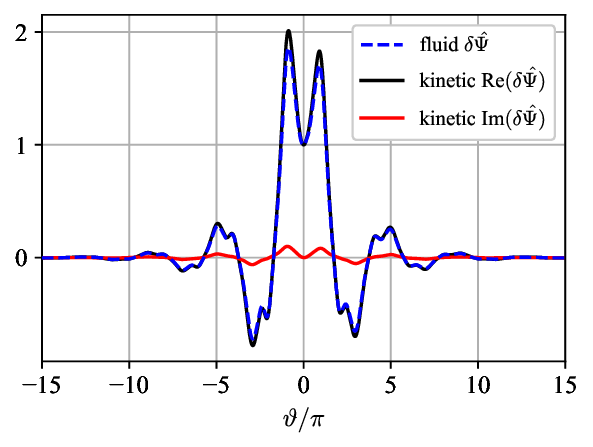}
  \caption{Parallel mode structures of TAE in fluid limit (blue dashed line) and kinetic description (black and red lines). }
  \label{mode_structure}
\end{figure}

As a simple application of the code, we use the DTT reference equilibrium and EP distribution function discussed above and obtain the TAE frequency $\Omega=0.584+0.00636i$, with $\Omega\equiv\omega R_0/v_{A0}$ and $v_{A0}$ the Alfvén speed on the magnetic axis. The parallel mode structure of TAE in the ballooning space is shown in figure \ref{mode_structure}. As a comparison, in the fluid limit, the upper and lower accumulation points of the TAE gap are given by $\Omega_U=0.642$ and $\Omega_L=0.334$, the TAE frequency locates at $\Omega=0.565$, and the corresponding parallel mode structure (blue dashed line in figure \ref{mode_structure}) closely resembles that of the kinetic results. These results suggest that the correction of kinetic effects can be regarded as perturbative for the present case. Perturbative analysis using equation (\ref{perturbative_eq}) indicates that the Landau damping rates resulting from thermal ions and electrons are $5.7\times 10^{-5}$ and $1.4\times 10^{-4}$, respectively, both of which are much smaller than the EP induced growth rate. The particularly small electron Landau damping on TAE is primarily attributed to the relatively high electron density and temperature in the scenario analyzed here, which prevents the efficient resonance between TAE and electrons, as demonstrated in figure \ref{deltaWk_Te}(b) and (d). Since circulating thermal electrons has negligible contribution to both the real frequency and damping rate of TAE, it is no longer considered in the subsequent calculations. 

\section{Effect of triangularity on TAE stability}\label{Effect of triangularity on TAE stability}
As a further demonstration of the code applicability, as well as the continuation of our previous work\cite{GWeiPoP2024}, in this section we investigate the effect of triangularity on TAE stability. In reference \citenum{GWeiPoP2024}, it has been found that triangularity has little impact on TAE in the fluid limit, except a small downward shift of the frequency. However, the introduction of kinetic effects may make a significant difference due to the modification of equilibrium orbits. For the convenience of controlling and parametrically modifying equilibrium geometry, we make use of the local Miller equilibrium with shape of plasma flux surface taking the form of\cite{RMillerPoP1998} 
\begin{equation}
  \begin{aligned}
    &R=R_0+r\cos(\theta+\delta\sin\theta),\\
    &Z=A r\sin\theta,
  \end{aligned}
\end{equation}
where $A$ and $\delta$ represent the elongation and triangularity, respectively. The parameters required for constructing the Miller equilibrium are fitted from the original DTT equilibrium at $\rho_{tor}=0.55$ with the fitted triangularity being $\delta=0.091$. Then we change the value of $\delta$ while keeping other parameters fixed, to investigate the effect of triangularity.

\subsection{Characteristic frequencies in PT and NT}

\begin{figure}[htbp]
  \includegraphics[width=0.40\textwidth]{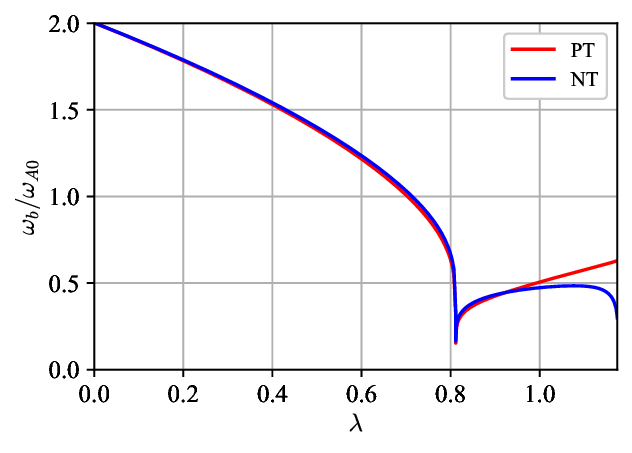}
  \caption{Dependence of transit/bounce frequency $\omega_b$ on $\lambda$ with $\mathcal{E}=\mathcal{E}_0$. }
  \label{omega_b}
\end{figure}

\begin{figure}[htbp]
  \includegraphics[width=0.40\textwidth]{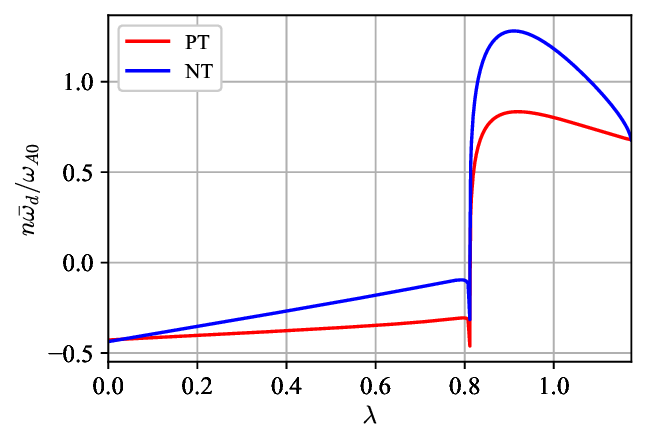}
  \caption{Dependence of precession frequency $n\bar{\omega}_d$ on $\lambda$ with $\mathcal{E}=\mathcal{E}_0$. Toroidal mode number $n$ is taken to be $10$ for reference. }
  \label{omega_d}
\end{figure}

First of all, we compare the particle characteristic equilibrium orbit frequencies as the function of $\lambda$ in positive triangularity (PT) with $\delta=0.091$ and negative triangularity (NT) with $\delta=-0.091$, as illustrated in figures \ref{omega_b} and \ref{omega_d}, which are calculated for fixed energy $\mathcal{E}=\mathcal{E}_0$, and $n=10$ is assumed for the evaluation of $n\bar{\omega}_d$. The energy dependences of $\omega_b$ and $\bar{\omega}_d$ are simply given by $\omega_b\propto\sqrt{\mathcal{E}}$ and $\bar{\omega}_d\propto\mathcal{E}$. Figures \ref{omega_b} and \ref{omega_d} indicate that switching the triangularity from positive to negative significantly modifies the precession frequencies of both circulating and trapped particles, as well as the bounce frequency of trapped particles, but has little impact on the transit frequency of circulating particle, which is consistent with the physical expectation: circulating particles experience the magnetic fields in the entire poloidal range, while trapped particles are mostly localized on the low field side, and are thus more strongly affected by the triangularity. Different from the transit frequency, the precession frequency is determined by the gradient of magnetic field, and is thus more sensitive to the geometry change. Notably, the enhancement of trapped particle precession frequency shown in figure \ref{omega_d} is consistent with the results in references \citenum{AMarinoniPPCF2009} and \citenum{JGravesPPCF2013}. Considering the significant differences in their characteristic frequencies, we discuss circulating or trapped EPs separately in the following study to better elucidate the different physical mechanisms that affect TAE stability. However, the kinetic contributions of all thermal species are included (except circulating electrons) in all the calculations regardless of their orbit types, as they primarily give rise to a frequency shift.

\subsection{Effect of triangularity on TAE stability via different mechanisms}

In this section, we discuss how triangularity can impact TAE stability via different underlying physics mechanisms. By simple inspection of equation (\ref{perturbative_eq}), it can be speculated that the triangularity can affect the TAE growth rate through the modification of three possible factors: (1) geometric coefficients, (2) resonance condition, and (3) mode frequency and mode structure. Generally speaking, these three factors are coupled together and should be determined self-consistently for a given plasma equilibrium. However, we will show subsequently that these factors have different relative importance on TAE growth rate under different circumstances. By studying them separately, we will be able to develop a deeper understanding of the corresponding phenomena and to demonstrate the general applicability of our results.

\begin{figure}[htbp]
  \includegraphics[width=0.40\textwidth]{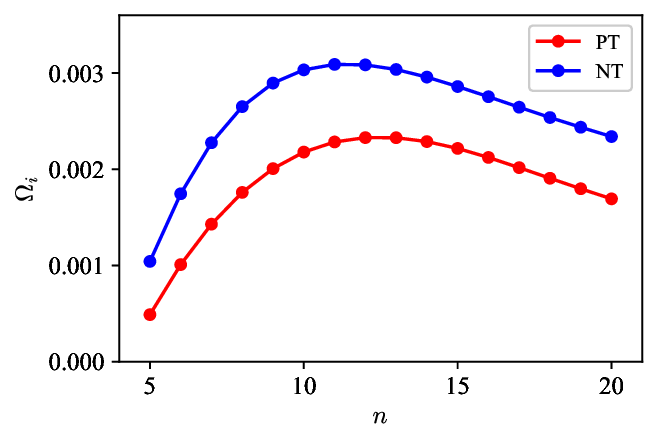}
  \caption{Growth rate of TAE versus toroidal mode number $n$ in PT (red circle) and NT (blue circle) when only circulating EPs are considered. }
  \label{n_scan_pass}
\end{figure}

\begin{figure}[htbp]
  \includegraphics[width=0.40\textwidth]{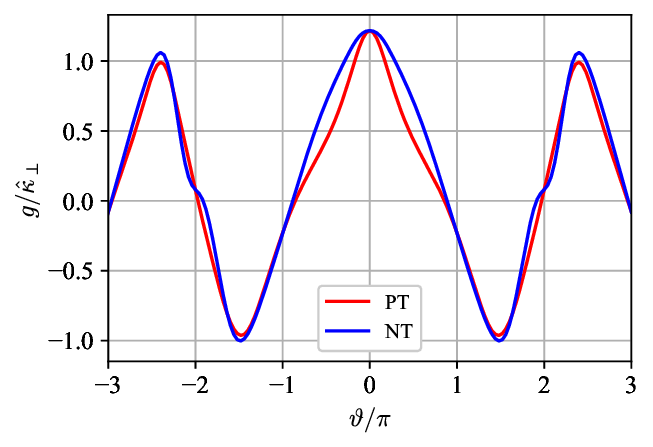}
  \caption{Coupling coefficient $g/\hat{\kappa}_\perp$ as a function of $\vartheta$ in the ballooning space in PT and NT. }
  \label{g_kappa_perp}
\end{figure}

\emph{Geometric coefficients.} Figure \ref{n_scan_pass} shows the TAE growth rate as a function of toroidal mode number when only circulating EPs are considered. In case with both positive and negative triangularities, the growth rates follow the same trend that first increases and then decreases, consistent with the extensive analytical and numerical studies\cite{GFuPoFB1992,LChenPoP1994,HVWongNF1995,FZoncaPoP1996}.
The TAE growth rate first increases with $n$ because the free energy associated with radial nonuniformity, accounted by the diamagnetic frequency $\omega_*$, is a linear function of $n$. Then it slowly decreases due to the averaging effect of FLR and FOW. For all the toroidal mode numbers listed in figure \ref{n_scan_pass}, TAE has larger growth rate in NT, which suggests a potential destabilization mechanisms in NT. 
In order to understand these results, we carry out a series of perturbative analyses using equation (\ref{perturbative_eq}) and including only EP kinetic compressibility, to isolate the effects of different factors. First of all, we calculate the $n=12$ TAE growth rates in both PT and NT using equation (\ref{perturbative_eq}), and the obtained results are $\Omega_i=2.6\times 10^{-3}$ and $\Omega_i=3.4\times 10^{-3}$, respectively, which approximately agree with the non-perturbative calculation results in figure \ref{n_scan_pass}, considering that core plasma compressibility is ignored here. Based on the perturbative calculation in PT, then we artificially replace the various elements in the expression of equation (\ref{perturbative_eq}) by those of NT and calculate TAE growth rate again, and the results are listed in table \ref{table1}. Replacing geometric coefficients leads to $\Omega_i=3.5\times 10^{-3}$, replacing mode frequency and mode structure leads to $\Omega_i=2.4\times 10^{-3}$, and finally, replacing particle orbit leads to $\Omega_i=2.7\times 10^{-3}$. Therefore, in this case, the change of geometric coefficients by the triangularity plays a dominant role in determining TAE growth rate. Further examinations identify that the coupling coefficient $g/\hat{\kappa}_\perp$ inside the kinetic compression term is the most significant one among various geometric coefficients, which is verified by the observation that TAE growth rates in PT and NT become very similar after imposing the same expression for $g/\hat{\kappa}_\perp$. In order to give a more intuitive physical picture, we plot $g/\hat{\kappa}_\perp$ as a function of $\vartheta$ in the ballooning space, as shown in figure \ref{g_kappa_perp}, which confirms the larger amplitude of $g/\hat{\kappa}_\perp$ in NT, especially in the small-$|\vartheta|$ region. Note that this coefficient appears in a `squared form' in the kinetic compression term, so the difference between PT and NT is actually more evident than what is shown here. From equation (\ref{g_func}), it can be recognized that $g/\hat{\kappa}_\perp$ is physically related to $\bm{v}_d\cdot\delta\bm{E}_\perp$, which controls the energy exchange between particles and wave. We also note that our result is consistent with the remark in reference \citenum{ABalestriPPCF2024} that magnetic drift velocity is faster in NT. It can also be anticipated that the Landau damping from core plasma could become stronger in NT for the same reason, but since its amplitude is always much smaller than EP drive for our parameters, it will not be discussed in more detail. The above mechanism for explaining the increased TAE growth rate in NT is valid not only for $n=12$, but also for the case with smaller $n$. However, as $n$ increases and becomes larger than 12, corresponding to the decreasing growth rate in figure \ref{n_scan_pass}, the perturbative analysis shows that the modification of particle orbit (and thus of the resonance condition) eventually becomes as important as the geometric coefficients at some point. This modification manifests primarily through the precession frequency shown in figure \ref{omega_d}, along with the strength of FLR and FOW effects $J_0e^{i\hat{Q}_B}$. For the case analyzed here, this additional mechanism also plays a role in destabilizing TAE in NT, but it is hard to discuss the generalizability of this result due to the complicated dependence of FLR and FOW effects on the triangularity.  

\begin{table}[htbp]
  \centering
  \caption{TAE growth rates obtained from self-consistent PT and NT calculations (first and second rows), and artificially replacing the various elements in the PT calculation by those of NT (last three rows). }
  \label{table1}
  \begin{tabular}{|c|c|}
    \hline
    Case & $\Omega_i$ \\
    \hline
    Self-consistent PT calculation & $2.6\times 10^{-3}$  \\
    Self-consistent NT calculation & $3.4\times 10^{-3}$  \\
    Replacing geometric coefficients  & $3.5\times 10^{-3}$  \\
    Replacing mode frequency and mode structure & $2.4\times 10^{-3}$  \\
    Replacing particle orbit & $2.7\times 10^{-3}$ \\
    \hline
  \end{tabular}
\end{table}

\begin{figure}[htbp]
  \includegraphics[width=0.40\textwidth]{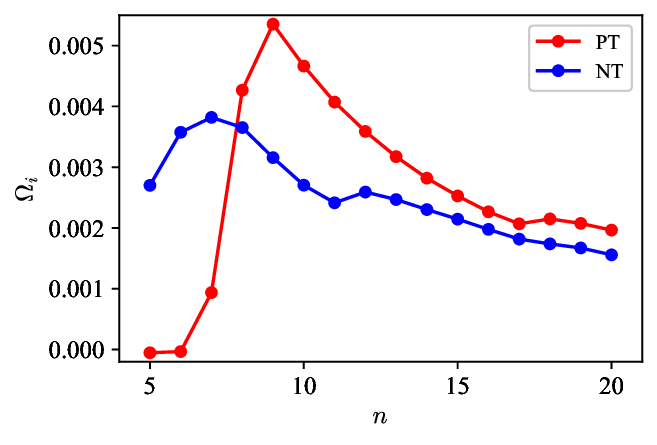}
  \caption{Growth rate of TAE versus toroidal mode number $n$ in PT (red circle) and NT (blue circle) when only trapped EPs are considered. }
  \label{n_scan_trap}
\end{figure}

\emph{Resonance condition.} Focusing now on the results for trapped EPs, figure \ref{n_scan_trap} indicates similar behavior of TAE growth rate with respect to $n$. However, the values of $n$ that maximize TAE instability are quite different in PT and NT, and TAE can be either more unstable or less unstable in NT, depending on the specific value of $n$. Before addressing the different results in the cases of PT and NT, it is useful to elucidate the underlying physics for a single curve in figure \ref{n_scan_trap}. Take the red curve for PT case as an example. In low-$n$ region ($n=5,6$), the mode is marginally stable due to the relatively small EP diamagnetic frequency and the absence of precession resonance. As $n$ increases, the rise in the diamagnetic frequency and the emergence of precession resonance $\omega=n\bar{\omega}_d$ leads to the rapid growth of TAE instability. Then for $n\geq 10$, the resonance lines (see figure \ref{deltaWk_EP}(d)), especially the $k=0$ one, gradually shift toward the lower energy region, which significantly reduces the strength of wave-particle resonance. Meanwhile, the stabilization effect of FLR and FOW also becomes important in the high-$n$ region. The combination of these factors results in a rapid decrease in TAE growth rate. At $n=18$, there appears another rise of the growth rate, mainly because the $k=-2$ resonance is triggered around $\mathcal{E}=\mathcal{E}_0$. Based on the understanding above, all the significant distinctions between the two curves in figure \ref{n_scan_trap} can be deduced from the change of precession frequency illustrated in figure \ref{omega_d}. In particular, due to the larger precession frequency in NT, the TAE growth rate reaches its peak, where precession resonance $\omega=n\bar{\omega}_d$ is optimal, at lower $n$ and consequently with lower $\omega_*$, as a result of which, the maximum TAE growth rate is correspondingly lower. These results indicate the sensitivity of resonance condition in determining TAE growth rate driven by trapped EPs. Finally, it is worthy noting that the relatively sharp inflection point observed in figure \ref{n_scan_trap} primarily stems from the truncation of slowing-down distribution function at the birth energy $\mathcal{E}_0$. By contrast, for EP with Maxwellian distribution, these curves are expected to have a smoother transition.

\begin{figure}[htbp]
  \includegraphics[width=0.40\textwidth]{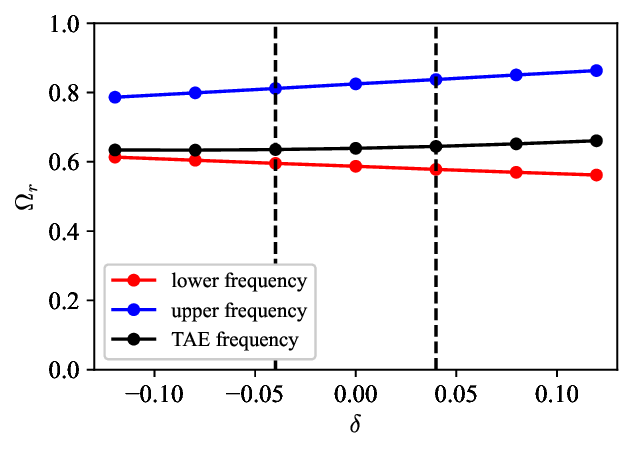}
  \caption{Real frequency of TAE (black circle) with $n=14$ versus the triangularity considering only circulating EPs. The blue and red circles represent the upper and lower accumulation points of TAE gap, respectively. The black dashed lines denote $\delta=\pm 0.04$.}
  \label{triangularity_scan_real}
\end{figure}

\begin{figure}[htbp]
  \includegraphics[width=0.40\textwidth]{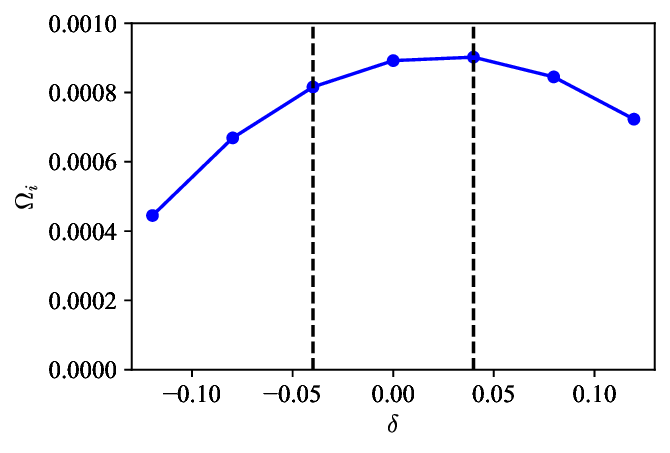}
  \caption{Growth rate of TAE with $n=14$ versus the triangularity considering only circulating EPs. }
  \label{triangularity_scan_imag}
\end{figure}

\emph{Mode frequency and mode structure.} In all the cases discussed above, we have ignored the change of mode frequency and mode structure for different triangularities, which can be justified provided that the mode is distant from the accumulation points. However, this assumption may not hold with the change of equilibrium parameters, such as magnetic shear and $\beta$ gradient\cite{FZoncaPoFB1993, GFuPoP1995}. In order to illustrate the importance of this mechanism, we use the local parameters of DTT at radial position $\rho_{tor}=0.32$ (see blue curve in figure \ref{equilibrium}(a)) to investigate how TAE stability is affected by the triangularity at smaller magnetic shear. Again, we make use of the local Miller equilibrium with fitted triangularity being $\delta=0.04$, and then change the value of $\delta$ while keeping other parameters fixed. Figures \ref{triangularity_scan_real} and \ref{triangularity_scan_imag} show the real frequency and growth rate of $n=14$ TAE versus the triangularity, where only circulating EPs are considered. The black dashed lines denote $\delta=\pm 0.04$. In figure \ref{triangularity_scan_real}, the positions of accumulation points are also depicted for reference, and it is observed that TAE frequency is relatively close to the lower accumulation point under this set of parameters. Besides, as triangularity decreases, the width of TAE gap shrinks gradually, and TAE real frequency slightly shifts downward, and consequently, approaches the lower accumulation point. Evidently, the EP contribution becomes non-perturbative in NT. Although the coupling coefficient $g/\hat{\kappa}_\perp$ is still enhanced in NT as shown in figure \ref{g_kappa_perp}, figure \ref{triangularity_scan_imag} indicates that TAE growth rate initially increases but subsequently decreases from PT to NT. Together with figure \ref{triangularity_scan_real}, this result demonstrates that the stronger coupling of TAE with SAW continuum leads to the stabilization of TAE in NT.

\section{Summary and prospect}\label{Summary and prospect}

Adopting the same theoretical framework and similar methodology as reference \citenum{GWeiPoP2024}, in this work a linear gyrokinetic eigenvalue code is developed to study TAE stability with the self-consistent treatment of EP drive and core plasma Landau damping in a non-perturbative way. The main motivation was to develop a numerical computation framework that could address drift Alfvén wave stability in general magnetic equilibria and with arbitrary particle distribution functions, deploying all the numerical diagnostics that are needed to analyze and interpret numerical results based on the well-established theoretical framework of the GFLDR\cite{LChenRMP2016,FZoncaPoP1996,FZoncaPRL1992}. The general particle responses of both circulating and trapped particles are calculated by either Fourier spectrum method or by direct integration along equilibrium particle orbits, employing an action-angle approach. In particular, the FLR and FOW effects of EPs are fully taken into account. The code supports different kinds of particle distribution functions, including Maxwellian, isotropic slowing-down and model anisotropic slowing-down distributions, which enables investigating the EP and AE physics in a broad range of applications of practical. Furthermore, the possibility of expansion to arbitrary distribution is also retained. The nonlinear eigenvalue problem is solved iteratively either by shooting method or finite element method, differing in the treatment of kinetic compression term. 
With the systematic implementation of numerical diagnostics, including the demonstration of effective mode structure and resonance structure, the code is able to study the physics of wave-particle interaction in great detail. For the time being, the ideal MHD approximation, i.e., vanishing parallel electric field, is adopted to simplify the eigenmode equations, which is justified for TAE generally dominated by Alfvénic polarization. 

As an application of the code, we perform an in-depth study of the triangularity effect on TAE stability adopting the reference equilibrium of DTT. Since our aim is to illuminate the physics processes underlying the dependence of TAE stability on equilibrium geometry, we adopt a simple isotropic alpha particle slowing down distribution to model the EP drive, discussing circulating and magnetically trapped particles separately. The effect of realistic particle sources will be addressed in a separate work. By constructing a series of local Miller equilibria, we explicitly isolate the effect of triangularity from other parameters. It is demonstrated that TAE growth rate can be affected by the triangularity through the modifications to geometric coupling coefficients, resonance condition, as well as the mode frequency and mode structure. The main results regarding triangularity effect are summarized as follows:
\begin{enumerate}
\item In NT, the wave-particle coupling coefficient $g/\hat{\kappa}_\perp$ is significantly enhanced, resulting in larger TAE growth rate. This mechanism is especially evident for circulating EPs, and is proved to be the dominant one when toroidal mode number is not very high, otherwise the modifications to FLR and FOW effects as well as the precession frequency should be taken into account properly.
\item The triangularity significantly modifies the resonance condition of trapped EPs through its impact on both precession and bounce frequencies. The stabilizing or destabilizing effect of NT on TAE depends critically on the toroidal mode number $n$. Nevertheless, it is found that due to the larger precession frequency of trapped particle in NT, the TAE growth rate reaches its peak at lower $n$ and consequently with lower diamagnetic frequency $\omega_*$. As a result, the maximum TAE growth rate is lower in NT plasmas. 
\item Scanning of triangularity reveals that the width of TAE gap shrinks gradually and TAE real frequency slightly shifts downward and approaches the lower accumulation point as the triangularity decreases. Therefore, for equilibrium with relatively low magnetic shear or strong $\beta$ gradient, TAE may be more strongly coupled with SAW continuum in NT plasmas and consequently less unstable.  
\end{enumerate}
Based on our studies, the overall effect of NT on TAE stability in a specific tokamak scenario can be evaluated and well understood. Moreover, it is quite obvious that NT does not generally bring improved stability or plasma performance.

With the non-perturbative treatment of particle kinetic effects in the calculation, the code is able to study not only TAE, but also the energetic particle mode (EPM)\cite{LChenPoP1994}, which arises from the SAW continuum and can be excited when EP drive is sufficiently strong and overcomes the continuum damping. However, for low frequency AEs such as BAE, our model will require further extension and incorporation of parallel electric field and thermal ion FLR effect. Furthermore, in the current work, the code is limited to the solution of local eigenvalue problem by adopting the ballooning-mode representation at the lowest order. The construction and solution of the self-consistent global eigenvalue problem will be addressed in our future work.

\section*{Acknowledgements}
This work was supported by National Key R\&D Program of China under Grant No. 2024YFE03170000, the Strategic Priority Research Program of Chinese Academy of Sciences under Grant No. XDB0790201, the National Science Foundation of China under Grant Nos. 12275236 and 12261131622, the Italian Ministry for Foreign Affairs and International Cooperation Project under Grant No. CN23GR02 and the MMNLP project CSN4 of INFN, Italy. 
This work has also partly been carried out within the framework of the EUROfusion Consortium, funded by the European Union via the Euratom Research and Training Programme (Grant Agreement No. 101052200-EUROfusion). 
Views and opinions expressed are, however, those of the author(s) only and do not necessarily reflect those of the European Union or the European Commission. Neither the European Union nor the European Commission can be held responsible for them.

\appendix
\section{Solutions of linear gyrokinetic equation by integration along the unperturbed orbits}\label{AppA}
In addition to the Fourier spectrum method introduced in section \ref{Solution method}, the linear gyrokinetic equation can also be solved by integration along the unperturbed orbits, proven to be equivalent. The derivation follows closely existing literature, but rather than solving $\delta\hat{K}$ directly, we first solve equation (\ref{GK_eq_drift_center}) in the drift/banana center frame to obtain $\delta\hat{K}_B$, then transform back to the guiding center frame to get $\delta\hat{K}$. 

For circulating particles\cite{PRutherfordPoF1968,JTaylorPP1968,WTangNF1980,GRewoldtPoFB1982,GFuPoFB1992,GFuPoFB1993,JKimPoFB1993,YLiPoP2020}, the solution of $\delta\hat{K}_B$ is given by 
\begin{equation}
  \begin{aligned}
  \delta\hat{K}_{B}(\theta_{c})&=i\frac{e}{m}\frac{QF_{0}}{\omega}\frac{1}{\omega_{b}}\int_{-\infty}^{\theta_{c}}d\theta_{c}'e^{i\hat{Q}_{B}'}\frac{J_{0}'\omega_d'}{\hat{\kappa}_{\perp}'}\delta\hat{\Psi}'\\
  &\times\exp\left[i\frac{\omega-n\bar{\omega}_{d}}{\omega_{b}}\left(\theta_{c}-\theta_{c}'\right)\right],
  \end{aligned}
\end{equation}
where we have made use of the boundary condition $\delta\hat{K}_B(\theta_c=-\infty)=0$. Again, circulating particles with both positive and negative $v_\parallel$ are taken into account through the mapping relation between $\theta_c$ and $\vartheta$.

For trapped particles\cite{PRutherfordPoF1968,JTaylorPP1968,WTangNF1980,LChenJGR1991,GRewoldtPoFB1982,GFuPoFB1992,GFuPoFB1993}, taking the closed bounce orbit in the range of $-\theta_{b}\le\vartheta\le\theta_{b}$ as an example, the solutions of equation (\ref{GK_eq_drift_center}) for particles with positive and negative $v_{\parallel}$, denoted as $\delta\hat{K}_B^+$ and $\delta\hat{K}_B^-$, are given by 
\begin{equation}
  \begin{aligned}
    \delta\hat{K}_{B}^{+}(\theta_{c})&=i\frac{e}{m}\frac{QF_{0}}{\omega}\frac{1}{\omega_{b}}\Bigg\{ \int_{-\pi/2}^{\theta_{c}}d\theta_{c}'e^{i\hat{Q}_{B}'}\frac{J_{0}'\omega_d'}{\hat{\kappa}_{\perp}'}\delta\hat{\Psi}'\\
    &\times\exp\left[i\frac{\omega-n\bar{\omega}_{d}}{\omega_{b}}\left(\theta_{c}-\theta_{c}'\right)\right]\\
    &+C_{1}\exp\left[i\frac{\omega-n\bar{\omega}_{d}}{\omega_{b}}\left(\theta_{c}+\pi/2\right)\right]\Bigg\} 
  \end{aligned}
\end{equation}
and 
\begin{equation}
  \begin{aligned}
    \delta\hat{K}_{B}^{-}(\theta_{c})&=i\frac{e}{m}\frac{QF_{0}}{\omega}\frac{1}{\omega_{b}}\Bigg\{ \int_{\pi/2}^{\theta_{c}}d\theta_{c}'e^{i\hat{Q}_{B}'}\frac{J_{0}'\omega_d'}{\hat{\kappa}_{\perp}'}\delta\hat{\Psi}'\\
    &\times\exp\left[i\frac{\omega-n\bar{\omega}_{d}}{\omega_{b}}\left(\theta_{c}-\theta_{c}'\right)\right]\\
    &+C_{2}\exp\left[i\frac{\omega-n\bar{\omega}_{d}}{\omega_{b}}\left(\theta_{c}-\pi/2\right)\right]\Bigg\}  
  \end{aligned}
\end{equation}
where $\delta\hat{K}_{B}^{+}(\theta_{c})$ and $\delta\hat{K}_{B}^{-}(\theta_{c})$ are defined in the range of $-\pi/2\le\theta_{c}\le\pi/2$, and $\pi/2\le\theta_{c}\le3\pi/2$, respectively. Integral constants $C_{1}$ and $C_{2}$ are determined by the boundary conditions, $\delta\hat{K}_B^{+}(\theta_c=-\pi/2)=\delta\hat{K}_B^{-}(\theta_c=3\pi/2)$ and $\delta\hat{K}_B^{+}(\theta_c=\pi/2)=\delta\hat{K}_B^{-}(\theta_c=\pi/2)$. 

The above solutions are implemented in our numerical code as well, and the corresponding results are very well consistent with the Fourier spectrum method.

\section{Solving the eigenvalue problem using finite element method}\label{AppB}
Following Refs.\citenum{YLiPoP2020} and \citenum{YLiPPCF2023}, using cubic B-spline as finite element basis functions, we represent the mode structure as $\delta\hat{\Psi}(\vartheta)=\sum_{j}c_{j}\phi_{j}(\vartheta)$, and construct the weak form of the vorticity equation 
\begin{equation}
  \begin{aligned}
    &\sum_{j}\phi_{i}\partial_{\vartheta}\phi_{j}\big|_{-\vartheta_{m}}^{\vartheta_{m}}c_{j}\\
    +&\sum_{j}\int_{-\vartheta_{m}}^{\vartheta_{m}}d\vartheta\big[-\partial_{\vartheta}\phi_{i}\partial_{\vartheta}\phi_{j}
    +\phi_{i}V_{f}(\omega;\vartheta)\phi_{j}\big]c_{j}\\
    =&\sum_{j}\int_{-\vartheta_{m}}^{\vartheta_{m}}d\vartheta\phi_{i}\text{KC}(\omega,\phi_{j};\vartheta)c_{j}.
  \end{aligned}
\end{equation}
Once the frequency in the kinetic compression term is given, this equation can be cast as a linear generalized eigenvalue problem, $\bm{A}\bm{x}=\omega^2\bm{B}\bm{x}$, where vector $\bm{x}$ is composed of the finite element coefficients $c_j$, matrix $\bm{B}$ corresponds to the inertia term on the left hand side of vorticity equation, and matrix $\bm{A}$ includes the contributions from all the other terms. Starting from a guess value of $\omega$, we can iteratively solve this problem until the solution of $\omega$ converges. Evidently, this approach has a better performance in terms of the convergence due to the consistent treatment of the mode structures on both sides of the vorticity equation. Therefore, it is believed to have a broader range of applications. However, this approach takes much longer computational time at each iteration than the one introduced in section \ref{Solution method}, because it requires the particle responses for all the basis functions rather than a single guess function (i.e., $\delta\hat{\Psi}_0$). 
Nevertheless, this approach is also implemented in our code, as a comparison and verification of the previous approach, and the relative differences of both mode frequency and growth rate obtained using the two approaches are generally less than 1\%.

\bibliographystyle{aipnum4-1}
\bibliography{ZQiubib}

\begin{thebibliography}{61}%
\makeatletter
\providecommand \@ifxundefined [1]{%
 \@ifx{#1\undefined}
}%
\providecommand \@ifnum [1]{%
 \ifnum #1\expandafter \@firstoftwo
 \else \expandafter \@secondoftwo
 \fi
}%
\providecommand \@ifx [1]{%
 \ifx #1\expandafter \@firstoftwo
 \else \expandafter \@secondoftwo
 \fi
}%
\providecommand \natexlab [1]{#1}%
\providecommand \enquote  [1]{``#1''}%
\providecommand \bibnamefont  [1]{#1}%
\providecommand \bibfnamefont [1]{#1}%
\providecommand \citenamefont [1]{#1}%
\providecommand \href@noop [0]{\@secondoftwo}%
\providecommand \href [0]{\begingroup \@sanitize@url \@href}%
\providecommand \@href[1]{\@@startlink{#1}\@@href}%
\providecommand \@@href[1]{\endgroup#1\@@endlink}%
\providecommand \@sanitize@url [0]{\catcode `\\12\catcode `\$12\catcode `\&12\catcode `\#12\catcode `\^12\catcode `\_12\catcode `\%12\relax}%
\providecommand \@@startlink[1]{}%
\providecommand \@@endlink[0]{}%
\providecommand \url  [0]{\begingroup\@sanitize@url \@url }%
\providecommand \@url [1]{\endgroup\@href {#1}{\urlprefix }}%
\providecommand \urlprefix  [0]{URL }%
\providecommand \Eprint [0]{\href }%
\providecommand \doibase [0]{http://dx.doi.org/}%
\providecommand \selectlanguage [0]{\@gobble}%
\providecommand \bibinfo  [0]{\@secondoftwo}%
\providecommand \bibfield  [0]{\@secondoftwo}%
\providecommand \translation [1]{[#1]}%
\providecommand \BibitemOpen [0]{}%
\providecommand \bibitemStop [0]{}%
\providecommand \bibitemNoStop [0]{.\EOS\space}%
\providecommand \EOS [0]{\spacefactor3000\relax}%
\providecommand \BibitemShut  [1]{\csname bibitem#1\endcsname}%
\let\auto@bib@innerbib\@empty
\bibitem [{\citenamefont {Hasegawa}\ and\ \citenamefont {Chen}(1974)}]{AHasegawaPRL1974}%
  \BibitemOpen
  \bibfield  {author} {\bibinfo {author} {\bibfnamefont {A.}~\bibnamefont {Hasegawa}}\ and\ \bibinfo {author} {\bibfnamefont {L.}~\bibnamefont {Chen}},\ }\href@noop {} {\bibfield  {journal} {\bibinfo  {journal} {Phys. Rev. Lett.}\ }\textbf {\bibinfo {volume} {32}},\ \bibinfo {pages} {454} (\bibinfo {year} {1974})}\BibitemShut {NoStop}%
\bibitem [{\citenamefont {Chen}\ and\ \citenamefont {Hasegawa}(1974)}]{LChenPoF1974}%
  \BibitemOpen
  \bibfield  {author} {\bibinfo {author} {\bibfnamefont {L.}~\bibnamefont {Chen}}\ and\ \bibinfo {author} {\bibfnamefont {A.}~\bibnamefont {Hasegawa}},\ }\href@noop {} {\bibfield  {journal} {\bibinfo  {journal} {The Physics of Fluids}\ }\textbf {\bibinfo {volume} {17}},\ \bibinfo {pages} {1399} (\bibinfo {year} {1974})}\BibitemShut {NoStop}%
\bibitem [{\citenamefont {Cheng}, \citenamefont {Chen},\ and\ \citenamefont {Chance}(1985)}]{CZChengAP1985}%
  \BibitemOpen
  \bibfield  {author} {\bibinfo {author} {\bibfnamefont {C.}~\bibnamefont {Cheng}}, \bibinfo {author} {\bibfnamefont {L.}~\bibnamefont {Chen}}, \ and\ \bibinfo {author} {\bibfnamefont {M.}~\bibnamefont {Chance}},\ }\href@noop {} {\bibfield  {journal} {\bibinfo  {journal} {Ann. Phys.}\ }\textbf {\bibinfo {volume} {161}},\ \bibinfo {pages} {21} (\bibinfo {year} {1985})}\BibitemShut {NoStop}%
\bibitem [{\citenamefont {Chu}\ \emph {et~al.}(1992)\citenamefont {Chu}, \citenamefont {Greene}, \citenamefont {Lao}, \citenamefont {Turnbull},\ and\ \citenamefont {Chance}}]{MSChuPoF1992}%
  \BibitemOpen
  \bibfield  {author} {\bibinfo {author} {\bibfnamefont {M.~S.}\ \bibnamefont {Chu}}, \bibinfo {author} {\bibfnamefont {J.~M.}\ \bibnamefont {Greene}}, \bibinfo {author} {\bibfnamefont {L.~L.}\ \bibnamefont {Lao}}, \bibinfo {author} {\bibfnamefont {A.~D.}\ \bibnamefont {Turnbull}}, \ and\ \bibinfo {author} {\bibfnamefont {M.~S.}\ \bibnamefont {Chance}},\ }\href {\doibase 10.1063/1.860327} {\bibfield  {journal} {\bibinfo  {journal} {Physics of Fluids B: Plasma Physics}\ }\textbf {\bibinfo {volume} {4}},\ \bibinfo {pages} {3713} (\bibinfo {year} {1992})}\BibitemShut {NoStop}%
\bibitem [{\citenamefont {Cheng}\ and\ \citenamefont {Chance}(1986)}]{CZChengPoF1986}%
  \BibitemOpen
  \bibfield  {author} {\bibinfo {author} {\bibfnamefont {C.~Z.}\ \bibnamefont {Cheng}}\ and\ \bibinfo {author} {\bibfnamefont {M.~S.}\ \bibnamefont {Chance}},\ }\href {\doibase 10.1063/1.865801} {\bibfield  {journal} {\bibinfo  {journal} {The Physics of Fluids}\ }\textbf {\bibinfo {volume} {29}},\ \bibinfo {pages} {3695} (\bibinfo {year} {1986})}\BibitemShut {NoStop}%
\bibitem [{\citenamefont {Heidbrink}\ \emph {et~al.}(1993)\citenamefont {Heidbrink}, \citenamefont {Strait}, \citenamefont {Chu},\ and\ \citenamefont {Turnbull}}]{WHeidbrinkPRL1993}%
  \BibitemOpen
  \bibfield  {author} {\bibinfo {author} {\bibfnamefont {W.}~\bibnamefont {Heidbrink}}, \bibinfo {author} {\bibfnamefont {E.}~\bibnamefont {Strait}}, \bibinfo {author} {\bibfnamefont {M.}~\bibnamefont {Chu}}, \ and\ \bibinfo {author} {\bibfnamefont {A.}~\bibnamefont {Turnbull}},\ }\href@noop {} {\bibfield  {journal} {\bibinfo  {journal} {Phys. Rev. Lett.}\ }\textbf {\bibinfo {volume} {71}},\ \bibinfo {pages} {855} (\bibinfo {year} {1993})}\BibitemShut {NoStop}%
\bibitem [{\citenamefont {Heidbrink}(2008)}]{WHeidbrinkPoP2008}%
  \BibitemOpen
  \bibfield  {author} {\bibinfo {author} {\bibfnamefont {W.~W.}\ \bibnamefont {Heidbrink}},\ }\href@noop {} {\bibfield  {journal} {\bibinfo  {journal} {Phys. Plasmas}\ }\textbf {\bibinfo {volume} {15}},\ \bibinfo {pages} {055501} (\bibinfo {year} {2008})}\BibitemShut {NoStop}%
\bibitem [{\citenamefont {Heidbrink}\ \emph {et~al.}(1991)\citenamefont {Heidbrink}, \citenamefont {Strait}, \citenamefont {Doyle}, \citenamefont {Sager},\ and\ \citenamefont {Snider}}]{WHeidbrinkPRL1991}%
  \BibitemOpen
  \bibfield  {author} {\bibinfo {author} {\bibfnamefont {W.}~\bibnamefont {Heidbrink}}, \bibinfo {author} {\bibfnamefont {E.}~\bibnamefont {Strait}}, \bibinfo {author} {\bibfnamefont {E.}~\bibnamefont {Doyle}}, \bibinfo {author} {\bibfnamefont {G.}~\bibnamefont {Sager}}, \ and\ \bibinfo {author} {\bibfnamefont {R.}~\bibnamefont {Snider}},\ }\href {\doibase 10.1088/0029-5515/31/9/002} {\bibfield  {journal} {\bibinfo  {journal} {Nuclear Fusion}\ }\textbf {\bibinfo {volume} {31}},\ \bibinfo {pages} {1635} (\bibinfo {year} {1991})}\BibitemShut {NoStop}%
\bibitem [{\citenamefont {Wong}\ \emph {et~al.}(1991)\citenamefont {Wong}, \citenamefont {Fonck}, \citenamefont {Paul}, \citenamefont {Roberts}, \citenamefont {Fredrickson}, \citenamefont {Nazikian}, \citenamefont {Park}, \citenamefont {Bell}, \citenamefont {Bretz}, \citenamefont {Budny}, \citenamefont {Cohen}, \citenamefont {Hammett}, \citenamefont {Jobes}, \citenamefont {Meade}, \citenamefont {Medley}, \citenamefont {Mueller}, \citenamefont {Nagayama}, \citenamefont {Owens},\ and\ \citenamefont {Synakowski}}]{KWongPRL1991}%
  \BibitemOpen
  \bibfield  {author} {\bibinfo {author} {\bibfnamefont {K.~L.}\ \bibnamefont {Wong}}, \bibinfo {author} {\bibfnamefont {R.~J.}\ \bibnamefont {Fonck}}, \bibinfo {author} {\bibfnamefont {S.~F.}\ \bibnamefont {Paul}}, \bibinfo {author} {\bibfnamefont {D.~R.}\ \bibnamefont {Roberts}}, \bibinfo {author} {\bibfnamefont {E.~D.}\ \bibnamefont {Fredrickson}}, \bibinfo {author} {\bibfnamefont {R.}~\bibnamefont {Nazikian}}, \bibinfo {author} {\bibfnamefont {H.~K.}\ \bibnamefont {Park}}, \bibinfo {author} {\bibfnamefont {M.}~\bibnamefont {Bell}}, \bibinfo {author} {\bibfnamefont {N.~L.}\ \bibnamefont {Bretz}}, \bibinfo {author} {\bibfnamefont {R.}~\bibnamefont {Budny}}, \bibinfo {author} {\bibfnamefont {S.}~\bibnamefont {Cohen}}, \bibinfo {author} {\bibfnamefont {G.~W.}\ \bibnamefont {Hammett}}, \bibinfo {author} {\bibfnamefont {F.~C.}\ \bibnamefont {Jobes}}, \bibinfo {author} {\bibfnamefont {D.~M.}\ \bibnamefont {Meade}}, \bibinfo {author} {\bibfnamefont {S.~S.}\ \bibnamefont {Medley}}, \bibinfo {author} {\bibfnamefont {D.}~\bibnamefont {Mueller}}, \bibinfo {author} {\bibfnamefont {Y.}~\bibnamefont {Nagayama}}, \bibinfo {author} {\bibfnamefont {D.~K.}\ \bibnamefont {Owens}}, \ and\ \bibinfo {author} {\bibfnamefont {E.~J.}\ \bibnamefont {Synakowski}},\ }\href@noop {} {\bibfield  {journal} {\bibinfo  {journal} {Phys. Rev. Lett.}\ }\textbf {\bibinfo {volume} {66}},\ \bibinfo {pages} {1874} (\bibinfo {year} {1991})}\BibitemShut {NoStop}%
\bibitem [{\citenamefont {Garc\'{\i}a-Mu\~noz}\ \emph {et~al.}(2010)\citenamefont {Garc\'{\i}a-Mu\~noz}, \citenamefont {Hicks}, \citenamefont {van Voornveld}, \citenamefont {Classen}, \citenamefont {Bilato}, \citenamefont {Bobkov}, \citenamefont {Bruedgam}, \citenamefont {Fahrbach}, \citenamefont {Igochine}, \citenamefont {Jaemsae}, \citenamefont {Maraschek},\ and\ \citenamefont {Sassenberg}}]{MGarciaMunozPRL2010}%
  \BibitemOpen
  \bibfield  {author} {\bibinfo {author} {\bibfnamefont {M.}~\bibnamefont {Garc\'{\i}a-Mu\~noz}}, \bibinfo {author} {\bibfnamefont {N.}~\bibnamefont {Hicks}}, \bibinfo {author} {\bibfnamefont {R.}~\bibnamefont {van Voornveld}}, \bibinfo {author} {\bibfnamefont {I.~G.~J.}\ \bibnamefont {Classen}}, \bibinfo {author} {\bibfnamefont {R.}~\bibnamefont {Bilato}}, \bibinfo {author} {\bibfnamefont {V.}~\bibnamefont {Bobkov}}, \bibinfo {author} {\bibfnamefont {M.}~\bibnamefont {Bruedgam}}, \bibinfo {author} {\bibfnamefont {H.-U.}\ \bibnamefont {Fahrbach}}, \bibinfo {author} {\bibfnamefont {V.}~\bibnamefont {Igochine}}, \bibinfo {author} {\bibfnamefont {S.}~\bibnamefont {Jaemsae}}, \bibinfo {author} {\bibfnamefont {M.}~\bibnamefont {Maraschek}}, \ and\ \bibinfo {author} {\bibfnamefont {K.}~\bibnamefont {Sassenberg}} (\bibinfo {collaboration} {ASDEX Upgrade Team}),\ }\href {\doibase 10.1103/PhysRevLett.104.185002} {\bibfield  {journal} {\bibinfo  {journal} {Phys. Rev. Lett.}\ }\textbf {\bibinfo {volume} {104}},\ \bibinfo {pages} {185002} (\bibinfo {year} {2010})}\BibitemShut {NoStop}%
\bibitem [{\citenamefont {Chen}\ and\ \citenamefont {Zonca}(2016)}]{LChenRMP2016}%
  \BibitemOpen
  \bibfield  {author} {\bibinfo {author} {\bibfnamefont {L.}~\bibnamefont {Chen}}\ and\ \bibinfo {author} {\bibfnamefont {F.}~\bibnamefont {Zonca}},\ }\href@noop {} {\bibfield  {journal} {\bibinfo  {journal} {Review of Modern Physics}\ }\textbf {\bibinfo {volume} {88}},\ \bibinfo {pages} {015008} (\bibinfo {year} {2016})}\BibitemShut {NoStop}%
\bibitem [{\citenamefont {Todo}(2019)}]{YTodoRMPP2019}%
  \BibitemOpen
  \bibfield  {author} {\bibinfo {author} {\bibfnamefont {Y.}~\bibnamefont {Todo}},\ }\href@noop {} {\bibfield  {journal} {\bibinfo  {journal} {Reviews of Modern Plasma Physics}\ }\textbf {\bibinfo {volume} {3}} (\bibinfo {year} {2019})}\BibitemShut {NoStop}%
\bibitem [{\citenamefont {Fu}\ and\ \citenamefont {Van~Dam}(1989)}]{GFuPoFB1989}%
  \BibitemOpen
  \bibfield  {author} {\bibinfo {author} {\bibfnamefont {G.~Y.}\ \bibnamefont {Fu}}\ and\ \bibinfo {author} {\bibfnamefont {J.~W.}\ \bibnamefont {Van~Dam}},\ }\href@noop {} {\bibfield  {journal} {\bibinfo  {journal} {Physics of Fluids B}\ }\textbf {\bibinfo {volume} {1}},\ \bibinfo {pages} {1949} (\bibinfo {year} {1989})}\BibitemShut {NoStop}%
\bibitem [{\citenamefont {Fu}\ and\ \citenamefont {Cheng}(1992)}]{GFuPoFB1992}%
  \BibitemOpen
  \bibfield  {author} {\bibinfo {author} {\bibfnamefont {G.~Y.}\ \bibnamefont {Fu}}\ and\ \bibinfo {author} {\bibfnamefont {C.~Z.}\ \bibnamefont {Cheng}},\ }\href {\doibase 10.1063/1.860328} {\bibfield  {journal} {\bibinfo  {journal} {Physics of Fluids B: Plasma Physics}\ }\textbf {\bibinfo {volume} {4}},\ \bibinfo {pages} {3722} (\bibinfo {year} {1992})}\BibitemShut {NoStop}%
\bibitem [{\citenamefont {Fu}, \citenamefont {Cheng},\ and\ \citenamefont {Wong}(1993)}]{GFuPoFB1993}%
  \BibitemOpen
  \bibfield  {author} {\bibinfo {author} {\bibfnamefont {G.~Y.}\ \bibnamefont {Fu}}, \bibinfo {author} {\bibfnamefont {C.~Z.}\ \bibnamefont {Cheng}}, \ and\ \bibinfo {author} {\bibfnamefont {K.~L.}\ \bibnamefont {Wong}},\ }\href {\doibase 10.1063/1.860572} {\bibfield  {journal} {\bibinfo  {journal} {Physics of Fluids B: Plasma Physics}\ }\textbf {\bibinfo {volume} {5}},\ \bibinfo {pages} {4040} (\bibinfo {year} {1993})}\BibitemShut {NoStop}%
\bibitem [{\citenamefont {Chen}(1994)}]{LChenPoP1994}%
  \BibitemOpen
  \bibfield  {author} {\bibinfo {author} {\bibfnamefont {L.}~\bibnamefont {Chen}},\ }\href@noop {} {\bibfield  {journal} {\bibinfo  {journal} {Physics of Plasmas}\ }\textbf {\bibinfo {volume} {1}},\ \bibinfo {pages} {1519} (\bibinfo {year} {1994})}\BibitemShut {NoStop}%
\bibitem [{\citenamefont {Zonca}\ and\ \citenamefont {Chen}(1996)}]{FZoncaPoP1996}%
  \BibitemOpen
  \bibfield  {author} {\bibinfo {author} {\bibfnamefont {F.}~\bibnamefont {Zonca}}\ and\ \bibinfo {author} {\bibfnamefont {L.}~\bibnamefont {Chen}},\ }\href@noop {} {\bibfield  {journal} {\bibinfo  {journal} {Physics of Plasmas}\ }\textbf {\bibinfo {volume} {3}},\ \bibinfo {pages} {323} (\bibinfo {year} {1996})}\BibitemShut {NoStop}%
\bibitem [{\citenamefont {Zonca}\ and\ \citenamefont {Chen}(1992)}]{FZoncaPRL1992}%
  \BibitemOpen
  \bibfield  {author} {\bibinfo {author} {\bibfnamefont {F.}~\bibnamefont {Zonca}}\ and\ \bibinfo {author} {\bibfnamefont {L.}~\bibnamefont {Chen}},\ }\href@noop {} {\bibfield  {journal} {\bibinfo  {journal} {Phys. Rev. Lett.}\ }\textbf {\bibinfo {volume} {68}},\ \bibinfo {pages} {592} (\bibinfo {year} {1992})}\BibitemShut {NoStop}%
\bibitem [{\citenamefont {Zonca}\ and\ \citenamefont {Chen}(1993)}]{FZoncaPoFB1993}%
  \BibitemOpen
  \bibfield  {author} {\bibinfo {author} {\bibfnamefont {F.}~\bibnamefont {Zonca}}\ and\ \bibinfo {author} {\bibfnamefont {L.}~\bibnamefont {Chen}},\ }\href@noop {} {\bibfield  {journal} {\bibinfo  {journal} {Physics of Fluids B: Plasma Physics}\ }\textbf {\bibinfo {volume} {5}},\ \bibinfo {pages} {3668} (\bibinfo {year} {1993})}\BibitemShut {NoStop}%
\bibitem [{\citenamefont {Berk}\ \emph {et~al.}(1992)\citenamefont {Berk}, \citenamefont {Van~Dam}, \citenamefont {Guo},\ and\ \citenamefont {Lindberg}}]{HBerkPoFB1992}%
  \BibitemOpen
  \bibfield  {author} {\bibinfo {author} {\bibfnamefont {H.~L.}\ \bibnamefont {Berk}}, \bibinfo {author} {\bibfnamefont {J.~W.}\ \bibnamefont {Van~Dam}}, \bibinfo {author} {\bibfnamefont {Z.}~\bibnamefont {Guo}}, \ and\ \bibinfo {author} {\bibfnamefont {D.~M.}\ \bibnamefont {Lindberg}},\ }\href {\doibase 10.1063/1.860455} {\bibfield  {journal} {\bibinfo  {journal} {Physics of Fluids B: Plasma Physics}\ }\textbf {\bibinfo {volume} {4}},\ \bibinfo {pages} {1806} (\bibinfo {year} {1992})}\BibitemShut {NoStop}%
\bibitem [{\citenamefont {Rosenbluth}\ \emph {et~al.}(1992)\citenamefont {Rosenbluth}, \citenamefont {Berk}, \citenamefont {Van~Dam},\ and\ \citenamefont {Lindberg}}]{MRosenbluthPRL1992}%
  \BibitemOpen
  \bibfield  {author} {\bibinfo {author} {\bibfnamefont {M.~N.}\ \bibnamefont {Rosenbluth}}, \bibinfo {author} {\bibfnamefont {H.~L.}\ \bibnamefont {Berk}}, \bibinfo {author} {\bibfnamefont {J.~W.}\ \bibnamefont {Van~Dam}}, \ and\ \bibinfo {author} {\bibfnamefont {D.~M.}\ \bibnamefont {Lindberg}},\ }\href@noop {} {\bibfield  {journal} {\bibinfo  {journal} {Phys. Rev. Lett.}\ }\textbf {\bibinfo {volume} {68}},\ \bibinfo {pages} {596} (\bibinfo {year} {1992})}\BibitemShut {NoStop}%
\bibitem [{\citenamefont {Mett}\ and\ \citenamefont {Mahajan}(1992)}]{RMettPoFB1992}%
  \BibitemOpen
  \bibfield  {author} {\bibinfo {author} {\bibfnamefont {R.~R.}\ \bibnamefont {Mett}}\ and\ \bibinfo {author} {\bibfnamefont {S.~M.}\ \bibnamefont {Mahajan}},\ }\href@noop {} {\bibfield  {journal} {\bibinfo  {journal} {Physics of Fluids B: Plasma Physics}\ }\textbf {\bibinfo {volume} {4}},\ \bibinfo {pages} {2885} (\bibinfo {year} {1992})}\BibitemShut {NoStop}%
\bibitem [{\citenamefont {Fu}\ \emph {et~al.}(1996)\citenamefont {Fu}, \citenamefont {Cheng}, \citenamefont {Budny}, \citenamefont {Chang}, \citenamefont {Darrow}, \citenamefont {Fredrickson}, \citenamefont {Mazzucato}, \citenamefont {Nazikian}, \citenamefont {Wong},\ and\ \citenamefont {Zweben}}]{GFuPoP1996}%
  \BibitemOpen
  \bibfield  {author} {\bibinfo {author} {\bibfnamefont {G.~Y.}\ \bibnamefont {Fu}}, \bibinfo {author} {\bibfnamefont {C.~Z.}\ \bibnamefont {Cheng}}, \bibinfo {author} {\bibfnamefont {R.}~\bibnamefont {Budny}}, \bibinfo {author} {\bibfnamefont {Z.}~\bibnamefont {Chang}}, \bibinfo {author} {\bibfnamefont {D.~S.}\ \bibnamefont {Darrow}}, \bibinfo {author} {\bibfnamefont {E.}~\bibnamefont {Fredrickson}}, \bibinfo {author} {\bibfnamefont {E.}~\bibnamefont {Mazzucato}}, \bibinfo {author} {\bibfnamefont {R.}~\bibnamefont {Nazikian}}, \bibinfo {author} {\bibfnamefont {K.~L.}\ \bibnamefont {Wong}}, \ and\ \bibinfo {author} {\bibfnamefont {S.}~\bibnamefont {Zweben}},\ }\href {\doibase 10.1063/1.871537} {\bibfield  {journal} {\bibinfo  {journal} {Physics of Plasmas}\ }\textbf {\bibinfo {volume} {3}},\ \bibinfo {pages} {4036} (\bibinfo {year} {1996})}\BibitemShut {NoStop}%
\bibitem [{\citenamefont {Fu}\ and\ \citenamefont {Berk}(2006)}]{GFuPoP2006}%
  \BibitemOpen
  \bibfield  {author} {\bibinfo {author} {\bibfnamefont {G.~Y.}\ \bibnamefont {Fu}}\ and\ \bibinfo {author} {\bibfnamefont {H.~L.}\ \bibnamefont {Berk}},\ }\href@noop {} {\bibfield  {journal} {\bibinfo  {journal} {Phys. Plasmas}\ }\textbf {\bibinfo {volume} {13}},\ \bibinfo {pages} {052502} (\bibinfo {year} {2006})}\BibitemShut {NoStop}%
\bibitem [{\citenamefont {Zonca}\ and\ \citenamefont {Chen}(2014{\natexlab{a}})}]{FZoncaPoP2014a}%
  \BibitemOpen
  \bibfield  {author} {\bibinfo {author} {\bibfnamefont {F.}~\bibnamefont {Zonca}}\ and\ \bibinfo {author} {\bibfnamefont {L.}~\bibnamefont {Chen}},\ }\href {\doibase 10.1063/1.4889019} {\bibfield  {journal} {\bibinfo  {journal} {Physics of Plasmas}\ }\textbf {\bibinfo {volume} {21}},\ \bibinfo {pages} {072120} (\bibinfo {year} {2014}{\natexlab{a}})}\BibitemShut {NoStop}%
\bibitem [{\citenamefont {Zonca}\ and\ \citenamefont {Chen}(2014{\natexlab{b}})}]{FZoncaPoP2014b}%
  \BibitemOpen
  \bibfield  {author} {\bibinfo {author} {\bibfnamefont {F.}~\bibnamefont {Zonca}}\ and\ \bibinfo {author} {\bibfnamefont {L.}~\bibnamefont {Chen}},\ }\href@noop {} {\bibfield  {journal} {\bibinfo  {journal} {Physics of Plasmas}\ }\textbf {\bibinfo {volume} {21}},\ \bibinfo {pages} {072121} (\bibinfo {year} {2014}{\natexlab{b}})}\BibitemShut {NoStop}%
\bibitem [{\citenamefont {Cheng}(1992)}]{CZChengPR1992}%
  \BibitemOpen
  \bibfield  {author} {\bibinfo {author} {\bibfnamefont {C.~Z.}\ \bibnamefont {Cheng}},\ }\href {\doibase 10.1016/0370-1573(92)90166-W} {\bibfield  {journal} {\bibinfo  {journal} {Physics Reports}\ }\textbf {\bibinfo {volume} {211}},\ \bibinfo {pages} {1} (\bibinfo {year} {1992})}\BibitemShut {NoStop}%
\bibitem [{\citenamefont {Borba}\ and\ \citenamefont {Kerner}(1999)}]{DBorbaJCP1999}%
  \BibitemOpen
  \bibfield  {author} {\bibinfo {author} {\bibfnamefont {D.}~\bibnamefont {Borba}}\ and\ \bibinfo {author} {\bibfnamefont {W.}~\bibnamefont {Kerner}},\ }\href {\doibase 10.1006/jcph.1999.6264} {\bibfield  {journal} {\bibinfo  {journal} {Journal of Computational Physics}\ }\textbf {\bibinfo {volume} {153}},\ \bibinfo {pages} {101} (\bibinfo {year} {1999})}\BibitemShut {NoStop}%
\bibitem [{\citenamefont {Lauber}\ \emph {et~al.}(2007)\citenamefont {Lauber}, \citenamefont {Günter}, \citenamefont {Könies},\ and\ \citenamefont {Pinches}}]{PLauberJCP2007}%
  \BibitemOpen
  \bibfield  {author} {\bibinfo {author} {\bibfnamefont {P.}~\bibnamefont {Lauber}}, \bibinfo {author} {\bibfnamefont {S.}~\bibnamefont {Günter}}, \bibinfo {author} {\bibfnamefont {A.}~\bibnamefont {Könies}}, \ and\ \bibinfo {author} {\bibfnamefont {S.~D.}\ \bibnamefont {Pinches}},\ }\href {\doibase 10.1016/j.jcp.2007.04.019} {\bibfield  {journal} {\bibinfo  {journal} {Journal of Computational Physics}\ }\textbf {\bibinfo {volume} {226}},\ \bibinfo {pages} {447} (\bibinfo {year} {2007})}\BibitemShut {NoStop}%
\bibitem [{\citenamefont {Bao}\ \emph {et~al.}(2023)\citenamefont {Bao}, \citenamefont {Zhang}, \citenamefont {Li}, \citenamefont {Lin}, \citenamefont {Dong}, \citenamefont {Liu}, \citenamefont {Xie}, \citenamefont {Meng}, \citenamefont {Cheng}, \citenamefont {Dong},\ and\ \citenamefont {Cao}}]{JBaoNF2023}%
  \BibitemOpen
  \bibfield  {author} {\bibinfo {author} {\bibfnamefont {J.}~\bibnamefont {Bao}}, \bibinfo {author} {\bibfnamefont {W.~L.}\ \bibnamefont {Zhang}}, \bibinfo {author} {\bibfnamefont {D.}~\bibnamefont {Li}}, \bibinfo {author} {\bibfnamefont {Z.}~\bibnamefont {Lin}}, \bibinfo {author} {\bibfnamefont {G.}~\bibnamefont {Dong}}, \bibinfo {author} {\bibfnamefont {C.}~\bibnamefont {Liu}}, \bibinfo {author} {\bibfnamefont {H.~S.}\ \bibnamefont {Xie}}, \bibinfo {author} {\bibfnamefont {G.}~\bibnamefont {Meng}}, \bibinfo {author} {\bibfnamefont {J.~Y.}\ \bibnamefont {Cheng}}, \bibinfo {author} {\bibfnamefont {C.}~\bibnamefont {Dong}}, \ and\ \bibinfo {author} {\bibfnamefont {J.~T.}\ \bibnamefont {Cao}},\ }\href {\doibase 10.1088/1741-4326/acd1a0} {\bibfield  {journal} {\bibinfo  {journal} {Nuclear Fusion}\ }\textbf {\bibinfo {volume} {63}},\ \bibinfo {pages} {076021} (\bibinfo {year} {2023})},\ \bibinfo {note} {publisher: IOP Publishing}\BibitemShut {NoStop}%
\bibitem [{\citenamefont {Wei}\ \emph {et~al.}(2024)\citenamefont {Wei}, \citenamefont {Falessi}, \citenamefont {Wang}, \citenamefont {Zonca},\ and\ \citenamefont {Qiu}}]{GWeiPoP2024}%
  \BibitemOpen
  \bibfield  {author} {\bibinfo {author} {\bibfnamefont {G.}~\bibnamefont {Wei}}, \bibinfo {author} {\bibfnamefont {M.~V.}\ \bibnamefont {Falessi}}, \bibinfo {author} {\bibfnamefont {T.}~\bibnamefont {Wang}}, \bibinfo {author} {\bibfnamefont {F.}~\bibnamefont {Zonca}}, \ and\ \bibinfo {author} {\bibfnamefont {Z.}~\bibnamefont {Qiu}},\ }\href {\doibase 10.1063/5.0213242} {\bibfield  {journal} {\bibinfo  {journal} {Physics of Plasmas}\ }\textbf {\bibinfo {volume} {31}},\ \bibinfo {pages} {072505} (\bibinfo {year} {2024})}\BibitemShut {NoStop}%
\bibitem [{\citenamefont {Zonca}\ \emph {et~al.}(2015)\citenamefont {Zonca}, \citenamefont {Chen}, \citenamefont {Briguglio}, \citenamefont {Fogaccia}, \citenamefont {Vlad},\ and\ \citenamefont {Wang}}]{FZoncaNJP2015}%
  \BibitemOpen
  \bibfield  {author} {\bibinfo {author} {\bibfnamefont {F.}~\bibnamefont {Zonca}}, \bibinfo {author} {\bibfnamefont {L.}~\bibnamefont {Chen}}, \bibinfo {author} {\bibfnamefont {S.}~\bibnamefont {Briguglio}}, \bibinfo {author} {\bibfnamefont {G.}~\bibnamefont {Fogaccia}}, \bibinfo {author} {\bibfnamefont {G.}~\bibnamefont {Vlad}}, \ and\ \bibinfo {author} {\bibfnamefont {X.}~\bibnamefont {Wang}},\ }\href@noop {} {\bibfield  {journal} {\bibinfo  {journal} {New Journal of Physics}\ }\textbf {\bibinfo {volume} {17}},\ \bibinfo {pages} {013052} (\bibinfo {year} {2015})}\BibitemShut {NoStop}%
\bibitem [{\citenamefont {Frieman}\ and\ \citenamefont {Chen}(1982)}]{EFriemanPoF1982}%
  \BibitemOpen
  \bibfield  {author} {\bibinfo {author} {\bibfnamefont {E.~A.}\ \bibnamefont {Frieman}}\ and\ \bibinfo {author} {\bibfnamefont {L.}~\bibnamefont {Chen}},\ }\href@noop {} {\bibfield  {journal} {\bibinfo  {journal} {Physics of Fluids}\ }\textbf {\bibinfo {volume} {25}},\ \bibinfo {pages} {502} (\bibinfo {year} {1982})}\BibitemShut {NoStop}%
\bibitem [{\citenamefont {Chen}\ and\ \citenamefont {Hasegawa}(1991)}]{LChenJGR1991}%
  \BibitemOpen
  \bibfield  {author} {\bibinfo {author} {\bibfnamefont {L.}~\bibnamefont {Chen}}\ and\ \bibinfo {author} {\bibfnamefont {A.}~\bibnamefont {Hasegawa}},\ }\href@noop {} {\bibfield  {journal} {\bibinfo  {journal} {Journal of Geophysical Research: Space Physics}\ }\textbf {\bibinfo {volume} {96}},\ \bibinfo {pages} {1503} (\bibinfo {year} {1991})}\BibitemShut {NoStop}%
\bibitem [{\citenamefont {Zonca}\ and\ \citenamefont {Chen}(2006)}]{FZoncaPPCF2006}%
  \BibitemOpen
  \bibfield  {author} {\bibinfo {author} {\bibfnamefont {F.}~\bibnamefont {Zonca}}\ and\ \bibinfo {author} {\bibfnamefont {L.}~\bibnamefont {Chen}},\ }\href {\doibase 10.1088/0741-3335/48/5/004} {\bibfield  {journal} {\bibinfo  {journal} {Plasma Physics and Controlled Fusion}\ }\textbf {\bibinfo {volume} {48}},\ \bibinfo {pages} {537} (\bibinfo {year} {2006})}\BibitemShut {NoStop}%
\bibitem [{\citenamefont {Connor}, \citenamefont {Hastie},\ and\ \citenamefont {Taylor}(1978)}]{JConnorPRL1978}%
  \BibitemOpen
  \bibfield  {author} {\bibinfo {author} {\bibfnamefont {J.}~\bibnamefont {Connor}}, \bibinfo {author} {\bibfnamefont {R.}~\bibnamefont {Hastie}}, \ and\ \bibinfo {author} {\bibfnamefont {J.}~\bibnamefont {Taylor}},\ }\href@noop {} {\bibfield  {journal} {\bibinfo  {journal} {Phys. Rev. Lett.}\ }\textbf {\bibinfo {volume} {40}},\ \bibinfo {pages} {396} (\bibinfo {year} {1978})}\BibitemShut {NoStop}%
\bibitem [{\citenamefont {Dewar}\ and\ \citenamefont {Glasser}(1983)}]{RDewarPoF1983}%
  \BibitemOpen
  \bibfield  {author} {\bibinfo {author} {\bibfnamefont {R.~L.}\ \bibnamefont {Dewar}}\ and\ \bibinfo {author} {\bibfnamefont {A.~H.}\ \bibnamefont {Glasser}},\ }\href {\doibase 10.1063/1.864028} {\bibfield  {journal} {\bibinfo  {journal} {The Physics of Fluids}\ }\textbf {\bibinfo {volume} {26}},\ \bibinfo {pages} {3038} (\bibinfo {year} {1983})}\BibitemShut {NoStop}%
\bibitem [{\citenamefont {Zonca}\ \emph {et~al.}(1999)\citenamefont {Zonca}, \citenamefont {Chen}, \citenamefont {Dong},\ and\ \citenamefont {Santoro}}]{FZoncaPoP1999}%
  \BibitemOpen
  \bibfield  {author} {\bibinfo {author} {\bibfnamefont {F.}~\bibnamefont {Zonca}}, \bibinfo {author} {\bibfnamefont {L.}~\bibnamefont {Chen}}, \bibinfo {author} {\bibfnamefont {J.~Q.}\ \bibnamefont {Dong}}, \ and\ \bibinfo {author} {\bibfnamefont {R.~A.}\ \bibnamefont {Santoro}},\ }\href@noop {} {\bibfield  {journal} {\bibinfo  {journal} {Physics of Plasmas}\ }\textbf {\bibinfo {volume} {6}},\ \bibinfo {pages} {1917} (\bibinfo {year} {1999})}\BibitemShut {NoStop}%
\bibitem [{\citenamefont {Brizard}\ and\ \citenamefont {Hahm}(2007)}]{ABrizardRMP2007}%
  \BibitemOpen
  \bibfield  {author} {\bibinfo {author} {\bibfnamefont {A.~J.}\ \bibnamefont {Brizard}}\ and\ \bibinfo {author} {\bibfnamefont {T.~S.}\ \bibnamefont {Hahm}},\ }\href@noop {} {\bibfield  {journal} {\bibinfo  {journal} {Rev. Mod. Phys.}\ }\textbf {\bibinfo {volume} {79}},\ \bibinfo {pages} {421} (\bibinfo {year} {2007})}\BibitemShut {NoStop}%
\bibitem [{\citenamefont {Cary}\ and\ \citenamefont {Brizard}(2009)}]{JcaryRMP2009}%
  \BibitemOpen
  \bibfield  {author} {\bibinfo {author} {\bibfnamefont {J.~R.}\ \bibnamefont {Cary}}\ and\ \bibinfo {author} {\bibfnamefont {A.~J.}\ \bibnamefont {Brizard}},\ }\href {\doibase 10.1103/RevModPhys.81.693} {\bibfield  {journal} {\bibinfo  {journal} {Reviews of Modern Physics}\ }\textbf {\bibinfo {volume} {81}},\ \bibinfo {pages} {693} (\bibinfo {year} {2009})},\ \bibinfo {note} {publisher: American Physical Society}\BibitemShut {NoStop}%
\bibitem [{\citenamefont {Tang}, \citenamefont {Connor},\ and\ \citenamefont {Hastie}(1980)}]{WTangNF1980}%
  \BibitemOpen
  \bibfield  {author} {\bibinfo {author} {\bibfnamefont {W.}~\bibnamefont {Tang}}, \bibinfo {author} {\bibfnamefont {J.}~\bibnamefont {Connor}}, \ and\ \bibinfo {author} {\bibfnamefont {R.}~\bibnamefont {Hastie}},\ }\href {\doibase 10.1088/0029-5515/20/11/011} {\bibfield  {journal} {\bibinfo  {journal} {Nuclear Fusion}\ }\textbf {\bibinfo {volume} {20}},\ \bibinfo {pages} {1439} (\bibinfo {year} {1980})}\BibitemShut {NoStop}%
\bibitem [{\citenamefont {Rewoldt}, \citenamefont {Tang},\ and\ \citenamefont {Chance}(1982)}]{GRewoldtPoFB1982}%
  \BibitemOpen
  \bibfield  {author} {\bibinfo {author} {\bibfnamefont {G.}~\bibnamefont {Rewoldt}}, \bibinfo {author} {\bibfnamefont {W.~M.}\ \bibnamefont {Tang}}, \ and\ \bibinfo {author} {\bibfnamefont {M.~S.}\ \bibnamefont {Chance}},\ }\href {\doibase 10.1063/1.863760} {\bibfield  {journal} {\bibinfo  {journal} {The Physics of Fluids}\ }\textbf {\bibinfo {volume} {25}},\ \bibinfo {pages} {480} (\bibinfo {year} {1982})}\BibitemShut {NoStop}%
\bibitem [{\citenamefont {Li}, \citenamefont {Hu},\ and\ \citenamefont {Zheng}(2020)}]{YLiPoP2020}%
  \BibitemOpen
  \bibfield  {author} {\bibinfo {author} {\bibfnamefont {Y.}~\bibnamefont {Li}}, \bibinfo {author} {\bibfnamefont {S.}~\bibnamefont {Hu}}, \ and\ \bibinfo {author} {\bibfnamefont {W.}~\bibnamefont {Zheng}},\ }\href@noop {} {\bibfield  {journal} {\bibinfo  {journal} {Physics of Plasmas}\ ,\ \bibinfo {pages} {12}} (\bibinfo {year} {2020})}\BibitemShut {NoStop}%
\bibitem [{\citenamefont {Wang}\ \emph {et~al.}(2018)\citenamefont {Wang}, \citenamefont {Qiu}, \citenamefont {Zonca}, \citenamefont {Briguglio}, \citenamefont {Fogaccia}, \citenamefont {Vlad},\ and\ \citenamefont {Wang}}]{TWangPoP2018}%
  \BibitemOpen
  \bibfield  {author} {\bibinfo {author} {\bibfnamefont {T.}~\bibnamefont {Wang}}, \bibinfo {author} {\bibfnamefont {Z.}~\bibnamefont {Qiu}}, \bibinfo {author} {\bibfnamefont {F.}~\bibnamefont {Zonca}}, \bibinfo {author} {\bibfnamefont {S.}~\bibnamefont {Briguglio}}, \bibinfo {author} {\bibfnamefont {G.}~\bibnamefont {Fogaccia}}, \bibinfo {author} {\bibfnamefont {G.}~\bibnamefont {Vlad}}, \ and\ \bibinfo {author} {\bibfnamefont {X.}~\bibnamefont {Wang}},\ }\href@noop {} {\bibfield  {journal} {\bibinfo  {journal} {Physics of Plasmas}\ }\textbf {\bibinfo {volume} {25}},\ \bibinfo {pages} {062509} (\bibinfo {year} {2018})}\BibitemShut {NoStop}%
\bibitem [{\citenamefont {Betti}\ and\ \citenamefont {Freidberg}(1991)}]{RBettiPoFB1991}%
  \BibitemOpen
  \bibfield  {author} {\bibinfo {author} {\bibfnamefont {R.}~\bibnamefont {Betti}}\ and\ \bibinfo {author} {\bibfnamefont {J.~P.}\ \bibnamefont {Freidberg}},\ }\href {\doibase 10.1063/1.859655} {\bibfield  {journal} {\bibinfo  {journal} {Physics of Fluids B: Plasma Physics}\ }\textbf {\bibinfo {volume} {3}},\ \bibinfo {pages} {1865} (\bibinfo {year} {1991})}\BibitemShut {NoStop}%
\bibitem [{\citenamefont {Betti}\ and\ \citenamefont {Freidberg}(1992)}]{RBettiPoFB1992}%
  \BibitemOpen
  \bibfield  {author} {\bibinfo {author} {\bibfnamefont {R.}~\bibnamefont {Betti}}\ and\ \bibinfo {author} {\bibfnamefont {J.~P.}\ \bibnamefont {Freidberg}},\ }\href {\doibase 10.1063/1.860057} {\bibfield  {journal} {\bibinfo  {journal} {Physics of Fluids B: Plasma Physics}\ }\textbf {\bibinfo {volume} {4}},\ \bibinfo {pages} {1465} (\bibinfo {year} {1992})}\BibitemShut {NoStop}%
\bibitem [{\citenamefont {Albanese}\ \emph {et~al.}(2003)\citenamefont {Albanese}, \citenamefont {Calabrò}, \citenamefont {Mattei},\ and\ \citenamefont {Villone}}]{RAlbaneseFED2003}%
  \BibitemOpen
  \bibfield  {author} {\bibinfo {author} {\bibfnamefont {R.}~\bibnamefont {Albanese}}, \bibinfo {author} {\bibfnamefont {G.}~\bibnamefont {Calabrò}}, \bibinfo {author} {\bibfnamefont {M.}~\bibnamefont {Mattei}}, \ and\ \bibinfo {author} {\bibfnamefont {F.}~\bibnamefont {Villone}},\ }\href {\doibase 10.1016/S0920-3796(03)00285-0} {\bibfield  {journal} {\bibinfo  {journal} {Fusion Engineering and Design}\ }\textbf {\bibinfo {volume} {66-68}},\ \bibinfo {pages} {715} (\bibinfo {year} {2003})},\ \bibinfo {note} {22nd Symposium on Fusion Technology}\BibitemShut {NoStop}%
\bibitem [{\citenamefont {Lütjens}, \citenamefont {Bondeson},\ and\ \citenamefont {Sauter}(1996)}]{HLutjensCPC1996}%
  \BibitemOpen
  \bibfield  {author} {\bibinfo {author} {\bibfnamefont {H.}~\bibnamefont {Lütjens}}, \bibinfo {author} {\bibfnamefont {A.}~\bibnamefont {Bondeson}}, \ and\ \bibinfo {author} {\bibfnamefont {O.}~\bibnamefont {Sauter}},\ }\href {\doibase 10.1016/0010-4655(96)00046-X} {\bibfield  {journal} {\bibinfo  {journal} {Computer Physics Communications}\ }\textbf {\bibinfo {volume} {97}},\ \bibinfo {pages} {219} (\bibinfo {year} {1996})}\BibitemShut {NoStop}%
\bibitem [{\citenamefont {Falessi}\ \emph {et~al.}(2019)\citenamefont {Falessi}, \citenamefont {Carlevaro}, \citenamefont {Fusco}, \citenamefont {Vlad},\ and\ \citenamefont {Zonca}}]{MFalessiPoP2019_continuum}%
  \BibitemOpen
  \bibfield  {author} {\bibinfo {author} {\bibfnamefont {M.~V.}\ \bibnamefont {Falessi}}, \bibinfo {author} {\bibfnamefont {N.}~\bibnamefont {Carlevaro}}, \bibinfo {author} {\bibfnamefont {V.}~\bibnamefont {Fusco}}, \bibinfo {author} {\bibfnamefont {G.}~\bibnamefont {Vlad}}, \ and\ \bibinfo {author} {\bibfnamefont {F.}~\bibnamefont {Zonca}},\ }\href {\doibase 10.1063/1.5098982} {\bibfield  {journal} {\bibinfo  {journal} {Physics of Plasmas}\ }\textbf {\bibinfo {volume} {26}},\ \bibinfo {pages} {082502} (\bibinfo {year} {2019})}\BibitemShut {NoStop}%
\bibitem [{\citenamefont {Falessi}\ \emph {et~al.}(2020)\citenamefont {Falessi}, \citenamefont {Carlevaro}, \citenamefont {Fusco}, \citenamefont {Giovannozzi}, \citenamefont {Lauber}, \citenamefont {Vlad},\ and\ \citenamefont {Zonca}}]{MFalessiJPP2020}%
  \BibitemOpen
  \bibfield  {author} {\bibinfo {author} {\bibfnamefont {M.~V.}\ \bibnamefont {Falessi}}, \bibinfo {author} {\bibfnamefont {N.}~\bibnamefont {Carlevaro}}, \bibinfo {author} {\bibfnamefont {V.}~\bibnamefont {Fusco}}, \bibinfo {author} {\bibfnamefont {E.}~\bibnamefont {Giovannozzi}}, \bibinfo {author} {\bibfnamefont {P.}~\bibnamefont {Lauber}}, \bibinfo {author} {\bibfnamefont {G.}~\bibnamefont {Vlad}}, \ and\ \bibinfo {author} {\bibfnamefont {F.}~\bibnamefont {Zonca}},\ }\href {\doibase 10.1017/S0022377820000975} {\bibfield  {journal} {\bibinfo  {journal} {Journal of Plasma Physics}\ }\textbf {\bibinfo {volume} {86}},\ \bibinfo {pages} {845860501} (\bibinfo {year} {2020})}\BibitemShut {NoStop}%
\bibitem [{\citenamefont {Stix}(1972)}]{TStixPoP1972}%
  \BibitemOpen
  \bibfield  {author} {\bibinfo {author} {\bibfnamefont {T.~H.}\ \bibnamefont {Stix}},\ }\href@noop {} {\bibfield  {journal} {\bibinfo  {journal} {Plasma Physics}\ }\textbf {\bibinfo {volume} {14}},\ \bibinfo {pages} {367} (\bibinfo {year} {1972})}\BibitemShut {NoStop}%
\bibitem [{\citenamefont {Miller}\ \emph {et~al.}(1998)\citenamefont {Miller}, \citenamefont {Chu}, \citenamefont {Greene}, \citenamefont {Lin-Liu},\ and\ \citenamefont {Waltz}}]{RMillerPoP1998}%
  \BibitemOpen
  \bibfield  {author} {\bibinfo {author} {\bibfnamefont {R.~L.}\ \bibnamefont {Miller}}, \bibinfo {author} {\bibfnamefont {M.~S.}\ \bibnamefont {Chu}}, \bibinfo {author} {\bibfnamefont {J.~M.}\ \bibnamefont {Greene}}, \bibinfo {author} {\bibfnamefont {Y.~R.}\ \bibnamefont {Lin-Liu}}, \ and\ \bibinfo {author} {\bibfnamefont {R.~E.}\ \bibnamefont {Waltz}},\ }\href {\doibase 10.1063/1.872666} {\bibfield  {journal} {\bibinfo  {journal} {Physics of Plasmas}\ }\textbf {\bibinfo {volume} {5}},\ \bibinfo {pages} {973} (\bibinfo {year} {1998})},\ \bibinfo {note} {publisher: American Institute of Physics}\BibitemShut {NoStop}%
\bibitem [{\citenamefont {Marinoni}\ \emph {et~al.}(2009)\citenamefont {Marinoni}, \citenamefont {Brunner}, \citenamefont {Camenen}, \citenamefont {Coda}, \citenamefont {Graves}, \citenamefont {Lapillonne}, \citenamefont {Pochelon}, \citenamefont {Sauter},\ and\ \citenamefont {Villard}}]{AMarinoniPPCF2009}%
  \BibitemOpen
  \bibfield  {author} {\bibinfo {author} {\bibfnamefont {A.}~\bibnamefont {Marinoni}}, \bibinfo {author} {\bibfnamefont {S.}~\bibnamefont {Brunner}}, \bibinfo {author} {\bibfnamefont {Y.}~\bibnamefont {Camenen}}, \bibinfo {author} {\bibfnamefont {S.}~\bibnamefont {Coda}}, \bibinfo {author} {\bibfnamefont {J.~P.}\ \bibnamefont {Graves}}, \bibinfo {author} {\bibfnamefont {X.}~\bibnamefont {Lapillonne}}, \bibinfo {author} {\bibfnamefont {A.}~\bibnamefont {Pochelon}}, \bibinfo {author} {\bibfnamefont {O.}~\bibnamefont {Sauter}}, \ and\ \bibinfo {author} {\bibfnamefont {L.}~\bibnamefont {Villard}},\ }\href {\doibase 10.1088/0741-3335/51/5/055016} {\bibfield  {journal} {\bibinfo  {journal} {Plasma Physics and Controlled Fusion}\ }\textbf {\bibinfo {volume} {51}},\ \bibinfo {pages} {055016} (\bibinfo {year} {2009})}\BibitemShut {NoStop}%
\bibitem [{\citenamefont {Graves}(2013)}]{JGravesPPCF2013}%
  \BibitemOpen
  \bibfield  {author} {\bibinfo {author} {\bibfnamefont {J.~P.}\ \bibnamefont {Graves}},\ }\href {\doibase 10.1088/0741-3335/55/7/074009} {\bibfield  {journal} {\bibinfo  {journal} {Plasma Physics and Controlled Fusion}\ }\textbf {\bibinfo {volume} {55}},\ \bibinfo {pages} {074009} (\bibinfo {year} {2013})}\BibitemShut {NoStop}%
\bibitem [{\citenamefont {Wong}, \citenamefont {Berk},\ and\ \citenamefont {Breizman}(1995)}]{HVWongNF1995}%
  \BibitemOpen
  \bibfield  {author} {\bibinfo {author} {\bibfnamefont {H.~V.}\ \bibnamefont {Wong}}, \bibinfo {author} {\bibfnamefont {H.~L.}\ \bibnamefont {Berk}}, \ and\ \bibinfo {author} {\bibfnamefont {B.~N.}\ \bibnamefont {Breizman}},\ }\href {\doibase 10.1088/0029-5515/35/12/I37} {\bibfield  {journal} {\bibinfo  {journal} {Nuclear Fusion}\ }\textbf {\bibinfo {volume} {35}},\ \bibinfo {pages} {1721} (\bibinfo {year} {1995})}\BibitemShut {NoStop}%
\bibitem [{\citenamefont {Balestri}\ \emph {et~al.}(2024)\citenamefont {Balestri}, \citenamefont {Ball}, \citenamefont {Coda}, \citenamefont {Cruz-Zabala}, \citenamefont {Garcia-Munoz},\ and\ \citenamefont {Viezzer}}]{ABalestriPPCF2024}%
  \BibitemOpen
  \bibfield  {author} {\bibinfo {author} {\bibfnamefont {A.}~\bibnamefont {Balestri}}, \bibinfo {author} {\bibfnamefont {J.}~\bibnamefont {Ball}}, \bibinfo {author} {\bibfnamefont {S.}~\bibnamefont {Coda}}, \bibinfo {author} {\bibfnamefont {D.~J.}\ \bibnamefont {Cruz-Zabala}}, \bibinfo {author} {\bibfnamefont {M.}~\bibnamefont {Garcia-Munoz}}, \ and\ \bibinfo {author} {\bibfnamefont {E.}~\bibnamefont {Viezzer}},\ }\href {\doibase 10.1088/1361-6587/ad4d1d} {\bibfield  {journal} {\bibinfo  {journal} {Plasma Physics and Controlled Fusion}\ }\textbf {\bibinfo {volume} {66}},\ \bibinfo {pages} {075012} (\bibinfo {year} {2024})}\BibitemShut {NoStop}%
\bibitem [{\citenamefont {Fu}(1995)}]{GFuPoP1995}%
  \BibitemOpen
  \bibfield  {author} {\bibinfo {author} {\bibfnamefont {G.~Y.}\ \bibnamefont {Fu}},\ }\href {\doibase 10.1063/1.871382} {\bibfield  {journal} {\bibinfo  {journal} {Physics of Plasmas}\ }\textbf {\bibinfo {volume} {2}},\ \bibinfo {pages} {1029} (\bibinfo {year} {1995})}\BibitemShut {NoStop}%
\bibitem [{\citenamefont {Rutherford}\ and\ \citenamefont {Frieman}(1968)}]{PRutherfordPoF1968}%
  \BibitemOpen
  \bibfield  {author} {\bibinfo {author} {\bibfnamefont {P.}~\bibnamefont {Rutherford}}\ and\ \bibinfo {author} {\bibfnamefont {E.}~\bibnamefont {Frieman}},\ }\href@noop {} {\bibfield  {journal} {\bibinfo  {journal} {Physics of Fluids}\ }\textbf {\bibinfo {volume} {11}},\ \bibinfo {pages} {569} (\bibinfo {year} {1968})}\BibitemShut {NoStop}%
\bibitem [{\citenamefont {Taylor}\ and\ \citenamefont {Hastie}(1968)}]{JTaylorPP1968}%
  \BibitemOpen
  \bibfield  {author} {\bibinfo {author} {\bibfnamefont {J.}~\bibnamefont {Taylor}}\ and\ \bibinfo {author} {\bibfnamefont {R.}~\bibnamefont {Hastie}},\ }\href@noop {} {\bibfield  {journal} {\bibinfo  {journal} {Plasma Physics}\ }\textbf {\bibinfo {volume} {10}},\ \bibinfo {pages} {479} (\bibinfo {year} {1968})}\BibitemShut {NoStop}%
\bibitem [{\citenamefont {Kim}, \citenamefont {Horton},\ and\ \citenamefont {Dong}(1993)}]{JKimPoFB1993}%
  \BibitemOpen
  \bibfield  {author} {\bibinfo {author} {\bibfnamefont {J.~Y.}\ \bibnamefont {Kim}}, \bibinfo {author} {\bibfnamefont {W.}~\bibnamefont {Horton}}, \ and\ \bibinfo {author} {\bibfnamefont {J.~Q.}\ \bibnamefont {Dong}},\ }\href@noop {} {\bibfield  {journal} {\bibinfo  {journal} {Physics of Fluids B: Plasma Physics}\ }\textbf {\bibinfo {volume} {5}},\ \bibinfo {pages} {4030} (\bibinfo {year} {1993})}\BibitemShut {NoStop}%
\bibitem [{\citenamefont {Li}\ \emph {et~al.}(2023)\citenamefont {Li}, \citenamefont {Falessi}, \citenamefont {Lauber}, \citenamefont {Li}, \citenamefont {Qiu}, \citenamefont {Wei},\ and\ \citenamefont {Zonca}}]{YLiPPCF2023}%
  \BibitemOpen
  \bibfield  {author} {\bibinfo {author} {\bibfnamefont {Y.}~\bibnamefont {Li}}, \bibinfo {author} {\bibfnamefont {M.~V.}\ \bibnamefont {Falessi}}, \bibinfo {author} {\bibfnamefont {P.}~\bibnamefont {Lauber}}, \bibinfo {author} {\bibfnamefont {Y.}~\bibnamefont {Li}}, \bibinfo {author} {\bibfnamefont {Z.}~\bibnamefont {Qiu}}, \bibinfo {author} {\bibfnamefont {G.}~\bibnamefont {Wei}}, \ and\ \bibinfo {author} {\bibfnamefont {F.}~\bibnamefont {Zonca}},\ }\href {\doibase 10.1088/1361-6587/acda5e} {\bibfield  {journal} {\bibinfo  {journal} {Plasma Physics and Controlled Fusion}\ }\textbf {\bibinfo {volume} {65}},\ \bibinfo {pages} {084001} (\bibinfo {year} {2023})}\BibitemShut {NoStop}%
\end{thebibliography}%

\end{document}